%% file: main.tex
\documentclass[%
            reprint,
            a4paper,
            superscriptaddress,
            amsmath,
            amssymb,
            aps,
            amsfonts,
            prmaterials,
            longbibliography,
            noshowpacs,
            floatfix,
]{revtex4-2}

\usepackage[pdftex]{hyperref}
\hypersetup{
	pdftitle={Lattice dynamics and complete polarization analysis of Raman-active modes in LaInO3},
	colorlinks,
	citecolor=blue,
	linkcolor=blue,
	urlcolor=blue
}

\usepackage[
,textwidth=17.5cm
,textheight=23.7cm
,verbose
,pdftex
]{geometry}

\usepackage[T1]{fontenc}
\usepackage{siunitx}
\usepackage{floatrow}
\usepackage[ragged,marginal,norule]{footmisc}
\usepackage{graphicx}
\usepackage{dcolumn}
\usepackage{bm}
\usepackage{multirow
            ,tikz
            ,paralist
            ,booktabs
            ,ulem
            ,soul
            ,bm
            }

\usepackage[percent]{overpic}
\usepackage{pgfplots}
\pgfplotsset{compat=1.18}
\usepackage{microtype}   
\usepackage{ragged2e}

\usepackage{color}

\newcommand{\rred}[1]{{\color{red}}}

\definecolor{darkgreen}{RGB}{0,128,0}

\newcommand{\aVar}{1}
\newcommand{\bVar}{0.7}
\newcommand{\cVar}{0.8}
\newcommand{\scaleVar}{0.25}
\newcommand{\tikzcenter}[1]{\vcenter{\hbox{#1}}}

\newcommand{\Beg}{$\rm{B_{1g}}$}
\newcommand{\Bzg}{$\rm{B_{2g}}$}
\newcommand{\Bdg}{$\rm{B_{3g}}$}
\newcommand{\Ag}{$\rm{A_{g}}$}
\newcommand{\mycomment}[1]{}

\DeclareSIUnit{\angstrom}{\textup{\AA}}

\renewcommand{\l}{\hspace{1ex}}

\begin{document}


\title{Lattice dynamics and complete polarization analysis of Raman-active modes in LaInO$_3$}

\author{Jonas Rose}
 \email{rose@pdi-berlin.de}
\affiliation{%
Paul-Drude-Institut für Festkörperelektronik, Leibniz-Institut im Forschungsverbund Berlin e.\,V., Hausvogteiplatz 5–7,
10117 Berlin, Germany.
}%
\author{Hai Nguyen}
\affiliation{%
Paul-Drude-Institut für Festkörperelektronik, Leibniz-Institut im Forschungsverbund Berlin e.\,V., Hausvogteiplatz 5–7,
10117 Berlin, Germany.
}%

\author{Moritz Mei\ss ner}
\affiliation{%
Paul-Drude-Institut für Festkörperelektronik, Leibniz-Institut im Forschungsverbund Berlin e.\,V., Hausvogteiplatz 5–7,
10117 Berlin, Germany.
}%

\author{Zbigniew Galazka}
\affiliation{%
 Leibniz-Institut für Kristallzüchtung, Max-Born-Str. 2, 12489 Berlin, Germany.
}%

\author{Roland Gillen}
\affiliation{%
Department of Electronic \& Electrical Engineering, Faculty of Science \& Engineering, Swansea University, Swansea, SA1 8EN, Wales, United Kingdom.
}%

\author{Georg Hoffmann}
\affiliation{%
Paul-Drude-Institut für Festkörperelektronik, Leibniz-Institut im Forschungsverbund Berlin e.\,V., Hausvogteiplatz 5–7,
10117 Berlin, Germany.
}%
\author{Oliver Brandt}
\affiliation{%
Paul-Drude-Institut für Festkörperelektronik, Leibniz-Institut im Forschungsverbund Berlin e.\,V., Hausvogteiplatz 5–7,
10117 Berlin, Germany.
}%
\author{Manfred Ramsteiner}
\affiliation{%
Paul-Drude-Institut für Festkörperelektronik, Leibniz-Institut im Forschungsverbund Berlin e.\,V., Hausvogteiplatz 5–7,
10117 Berlin, Germany.
}%
\author{Markus R. Wagner}
\affiliation{%
Paul-Drude-Institut für Festkörperelektronik, Leibniz-Institut im Forschungsverbund Berlin e.\,V., Hausvogteiplatz 5–7,
10117 Berlin, Germany.
}%

\author{Hans Tornatzky}
 \email{tornatzky@pdi-berlin.de}
\affiliation{%
Paul-Drude-Institut für Festkörperelektronik, Leibniz-Institut im Forschungsverbund Berlin e.\,V., Hausvogteiplatz 5–7,
10117 Berlin, Germany.
}%


\begin{abstract}
In this study, we present a comprehensive analysis of the Raman active phonon modes in orthorhombic LaInO$_3$ based on a combination of polarization-angle resolved Raman spectroscopy and density functional theory calculations. By using backscattering from multiple crystallographic surface orientations and employing a full symmetry analysis, we identify and assign most of the Raman-active $\Gamma$-point phonons to their irreducible representations of the D$_{\rm{2h}}$ point group. A multidimensional hyperspectral fitting procedure allows us to extract the relative Raman tensor elements from the angular dependence of the scattering intensities, even for strongly overlapping modes. First-principles calculations yield the phonon dispersion along high-symmetry directions, the phonon densities of states, and atomic displacement patterns, which are found to be in good agreement with the experimental mode frequencies.
\end{abstract}

\maketitle

\section{Introduction}

Over the last years, transparent conductive oxides (TCOs) have become a fundamental building block for new device concepts and an important topic of fundamental research.
In particular, perovskite oxides (ABO$_3$) exhibit multifunctional physical properties, spanning superconductivity, magnetism, and topological conductivity among others. Even more exciting, their ability to epitaxially stack them on top of each other generates new physical phenomena at their interfaces beyond their bulk properties, like two-dimensional (2D) superconductivity \cite{reyren_superconducting_2007}, 2D magnetism \cite{brinkman_magnetic_2007} or 2D electron gases (2DEG) \cite{ohtomo_high-mobility_2004}.\\
Within this family, La-doped BaSnO$_3$ (BSO) has emerged as a key material. Single crystal BSO exhibits the highest room-temperature electron mobility ($\mu \approx \SI{320}{\centi\metre\squared\per\volt\per\second}$ \cite{kim_high_2012}) reported for perovskite oxides while maintaining excellent optical transparency due to its wide band gap of \SI{3.1}{\electronvolt}. It further provides exceptional thermal stability \cite{kim_physical_2012} and a low oxygen diffusion coefficient \cite{belthle_quantitative_2022}. In epitaxial thin films, however, the mobility is typically reduced when using commercially available substrates such as SrTiO$_3$\cite{kim_all-perovskite_2015}, LaAlO$_3$\cite{kim_all-perovskite_2015}, MgO\cite{kim_laino3basno3_2018} or DyScO$_3$\cite{paik_adsorption-controlled_2017, hoffmann_enabling_2024}. This reduction is widely attributed to scattering at threading dislocations induced by substantial lattice mismatch of up to $5\%$ as well as crystallographic shear defects and point defects arising from nonstoichiometry \cite{paik_adsorption-controlled_2017, pfutzenreuter_epitaxial_2021}. Beyond advances in epitaxial growth, employing substrates with improved lattice compatibility therefore represents a promising strategy to enhance the electronic performance of BSO heterostructures. \\
To this end, LaInO$_3$ (LIO) has emerged as one of the most promising materials as it exhibits a pseudocubic lattice parameter that is nearly matched to that of BSO and can be grown from the melt \cite{galazka_melt_2021}. More interestingly, it was demonstrated that a $\rm{(SnO_{2})^0}$/$\rm{(LaO)^{+1}}$-terminated BSO/LIO interface gives rise to a 2DEG confined in the BSO through polar discontinuity doping \cite{kim_all-perovskite_2015, aggoune_tuning_2021}. Moreover, by employing LIO as a gate oxide, field effect transistors can be realized that offer high electron mobility, low sheet resistance and high charge carrier densities, positioning the BSO/LIO heterostructure as a compelling platform towards novel oxide electronic devices.\\
Despite the increasing relevance of LIO in this context, detailed investigations of its lattice dynamics remain scarce as previous studies suffered from inferior sample quality \cite{he_effects_2001, kumar_study_2017}. As phonon-related mechanisms are one of the most important foundations for understanding materials, covering mechanical and elastic properties, thermal transport as well as charge-carrier dynamics and phonon-assisted optical excitations, a comprehensive understanding of the lattice dynamics is essential for the proper design of associated processes in any application. \\

In this study, we investigate the lattice dynamics of bulk LIO by combining polarization-angle resolved Raman spectroscopy with density functional theory (DFT). Experimentally, we observe 19 of the 24 Raman-active modes and assign them to irreducible representations of the crystal's point group based on a symmetry analysis. By employing a multidimensional fitting procedure, we determine the relative Raman tensor elements of the observed modes from the angular dependencies of their corresponding measured scattering efficiencies.
Furthermore, we present the DFT-derived phonon dispersion along high-symmetry paths in reciprocal space, visualize the atomic displacement patterns for all $\Gamma$-point phonons and compare the calculated mode frequencies with our experimental observations.

\section{Crystal structure and Raman active phonons}
LIO crystallizes in an orthorhombic crystal structure and belongs to the space group $\rm{D^{16}_{2h}}$ in Schönflies notation (No.\ 62 in International Tables for Crystallography, ITC). There are two common Hermann-Mauguin notations for this space group, Pnma and Pbnm, which differ only by a cyclic permutation of the lattice vectors. To remain consistent with the ITC, all notations in this study follow the Pnma setting, in which the lattice constant $b_0$ (along $B$ axis) is the longest.
The crystal is fully described by a primitive orthorhombic Bravais lattice with lattice parameters of $a_0 = \SI{5.938}{\angstrom}$, $b_0 = \SI{8.214}{\angstrom}$ and $c_0 = \SI{5.723}{\angstrom}$ \cite{galazka_melt_2021} and a basis which is constructed using the Wyckoff sites and fractional coordinates given in Tab.\ \ref{tab:irrep}.

\begin{table*}[]
\caption{Wyckoff positions with site symmetry \cite{aroyo_bilbao_2006} and fractional coordinates of the atoms, calculated from DFT, and their contributions to the irreducible representations of the $\Gamma$-point phonons in Pnma space group \cite{kroumova_bilbao_2003}.}
\begin{tabular}{cccccc|cccc|ccc|c} \toprule \toprule
 & & & \multicolumn{3}{c|}{fractional coordinates} & \multicolumn{4}{c|}{Raman active} & \multicolumn{3}{c|}{\shortstack{IR active\\\& acoustic}} & silent \\
atom & site & symmetry & $x$ & $y$ & $z$ & \Ag{} & \Beg{} & \Bzg{} & \Bdg{} & $\rm{B_{1u}}$ & $\rm{B_{2u}}$ & $\rm{B_{3u}}$ & $\rm{A_u}$ \\ \midrule
La   & 4c & $\rm{C_{s}}$ & 0.9442 & 0.2500 & 0.0170 & 2 & 1 & 2 & 1 & 2 & 1 & 2 & 1 \\
In   & 4b & $\rm{C_{i}}$ & 0.0000 & 0.0000 & 0.5000 & -- & -- & -- & -- & 3 & 3 & 3 & 3 \\
O(1) & 4c & $\rm{C_{s}}$ & 0.0498 & 0.2500 & 0.6186 & 2 & 1 & 2 & 1 & 2 & 1 & 2 & 1 \\
O(2) & 8d & $\rm{C}_1$ & 0.1970 & 0.0628 & 0.1947 & 3 & 3 & 3 & 3 & 3 & 3 & 3 & 3 \\
\bottomrule \bottomrule
\label{tab:irrep}
\end{tabular}
\end{table*}

The atomic basis comprises 20 atoms ($Z=4$),
giving rise to 60 phonon modes. At the $\Gamma$-point, the corresponding atomic displacements decompose into the irreducible representations of the $\rm{D_{2h}}$ point group \cite{kroumova_bilbao_2003}:
\begin{align*}
\Gamma =& 7 \rm{A_{g}}\oplus5\rm{B_{1g}}\oplus7\rm{B_{2g}}\oplus5\rm{B_{3g}}\\
& \oplus10\rm{B_{1u}}\oplus8\rm{B_{2u}}\oplus10\rm{B_{3u}}\oplus8\rm{A_{u}} \text{.}
\end{align*}

\noindent Of these modes, 24 ($\rm{7 A_{g}\oplus5B_{1g}\oplus7B_{2g}\oplus5B_{3g}}$) are Raman active, 25 ($\rm{9B_{1u}\oplus7B_{2u}\oplus9B_{3u}}$) are IR active, 3 ($\rm{B_{1u}\oplus B_{2u}\oplus B_{3u}}$) are acoustic and 8 ($\rm{8A_u}$) are silent. Following Tab. \ref{tab:irrep}, In atoms do not contribute to Raman active modes as their site symmetry includes an inversion center, allowing participation only in odd-parity modes for the LIO crystal structure. 


Table \ref{tab:selection} summarizes the selection rules for all Raman-active phonon modes in the four different measurement geometries available to us. The Raman tensors $R$ are expressed in the crystal principal axis system ($A, B, C$), which is aligned with the fixed laboratory coordinate system ($X, Y, Z$). If a different axis is oriented along $Z$, i.\ e.\ if another surface plane is investigated, the transformed Raman tensors are obtained by applying the appropriate rotation matrices. In our experiments, the $Z$-axis is parallel to both the incident and scattered light, while the polarization is rotated within the $XY$-plane by an angle $\phi$. The corresponding polarization vectors are given by $\hat{e}_{\rm i}=(\cos(\phi), \sin(\phi),0)$ and $\hat{e}_{\rm s}=(\cos(\phi), \sin(\phi),0)$ for parallel or $\hat{e}_{\rm s}=(-\sin(\phi), \cos(\phi),0)$ for crossed polarization. Under these conditions, the third row and column of the transformed Raman tensors as well as the third entries of $\hat{e}_{\rm i,s}$ can be eliminated and the problem becomes two-dimensional.\\
Usually, the Raman intensity can be expressed as:

\begin{equation}
    I\propto \left| \hat{e}_{\rm i} R \hat{e}_{\rm s} \right|^2.
\label{eq:raman_int}
\end{equation}

\noindent Due to birefringence, i.\ e.\ different refractive indices along different axes, the portion of the sample above the scattering event can be considered as a waveplate, which leads to an additional depth-dependent rotation of $\hat{e}_{\rm i}$ and $\hat{e}_{\rm s}$. However, for excitations within the transparency regime of bulk crystals, where sufficiently large penetration depths are reached, and considerable birefringence, the depth-dependency can be neglected and the Raman intensity is calculated as \cite{kranert_raman_2016}

\begin{equation}
    I \propto \left| \hat{e}_{\rm i} R_0 \hat{e}_{\rm s} \right|^2 + \left| \hat{e}_{\rm i} R_1 \hat{e}_{\rm s} \right|^2 + \left| \hat{e}_{\rm i} R_2 \hat{e}_{\rm s} \right|^2,
\label{eq:intensity_biref}
\end{equation}

\noindent where

\begin{equation}
    R_0 = \begin{pmatrix} r_0 & 0\\0 & 0 \end{pmatrix}, \quad R_1=\begin{pmatrix} 0 & r_1\\r_1 & 0 \end{pmatrix}, \quad R_2=\begin{pmatrix} 0 & 0\\0 & r_2 \end{pmatrix}
\label{eq:Ramanmat}
\end{equation}

\noindent are summands of the transformed Raman tensor, such that $R=R_0+R_1+R_2$.\\
Note that birefringence has no influence on the $\rm{B_{xg}}$ modes in the geometries listed in Tab.\ \ref{tab:selection}. As their transformed Raman tensors only possess off-diagonal elements, $R_0$ and $R_2$ vanish and the expression in Eq.\ (\ref{eq:intensity_biref}) collapses to a single term identical to Eq.\ (\ref{eq:raman_int}).
Furthermore, birefringence affects the transmission of light at the air-sample interface, resulting in an additional angle-dependent modulation of the scattering intensity. This effect is taken into account by multiplying the transformed Raman tensor from left and right by the diagonal matrix $\mathrm{diag}(t_X/t_Y, 1)$\cite{kranert_raman_2016_GaO}, effectively scaling $r_0$ and $r_1$ from Eq.\ (\ref{eq:Ramanmat}) by $(t_X/t_Y)^2$ and $t_X/t_Y$, respectively. Here, $t_{\hat{e}} = 2/(n_{\hat{e}}+1)$ denotes the Fresnel transmission coefficient for light propagating along $Z$ and polarization along $\hat{e}$.\\
In order to compare the tensor elements of different modes with each other, a wavenumber-dependent prefactor \cite{HayesWilliam1978Solb}

\begin{equation}
    F(\omega_{\rm{P}}) = \frac{\omega_{\rm{L}}(\omega_{\rm{L}}-\omega_{\rm{P}})^3}{\omega_{\rm{P}}\left( 1-\exp\left(-\frac{\hbar\omega_{\rm{P}}}{\rm{k_B}\it{T}}\right)\right)}
\label{eq:prefactor}
\end{equation}

\noindent for the measured scattering intensity must be considered, where $\omega_{\rm{L}}$ is the laser and $\omega_{\rm{P}}$ the phonon angular frequency.

\begin{table*}[]
\caption{Selections rules of all Raman-active phonon modes in four different geometries including plots of their intensity profiles. The first row of each geometry corresponds to parallel (blue), the second row to crossed polarization (orange). The Raman tensors $R$ are expressed in the crystal principal axis  ($A$, $B$, $C$). $N$ denotes the surface normal of the (101) face. The ratios of the Raman tensor elements $a$, $b$, $c$ in the plotted \Ag{} mode intensity profiles are 10:7:8.}
\begin{tabular}{c<{\hspace{1em}}cc<{\hspace{1em}}cc<{\hspace{1em}}cc<{\hspace{1em}}cc}
\toprule \toprule
 & \multicolumn{2}{c}{\Ag{}} & \multicolumn{2}{c}{\Beg} & 
 \multicolumn{2}{c}{\Bzg} & \multicolumn{2}{c}{\Bdg} \\
 geometry & \multicolumn{2}{c}{$ \begin{pmatrix} a & 0 & 0\\ 0 & b & 0\\ 0 & 0 & c \end{pmatrix} $} & \multicolumn{2}{c}{$ \begin{pmatrix} 0 & d & 0\\ d & 0 & 0\\ 0 & 0 & 0 \end{pmatrix}$} & \multicolumn{2}{c}{$ \begin{pmatrix} 0 & 0 & e\\ 0 & 0 & 0\\ e & 0 & 0 \end{pmatrix} $} & \multicolumn{2}{c}{$ \begin{pmatrix} 0 & 0 & 0\\ 0 & 0 & f\\ 0 & f & 0 \end{pmatrix} $} \\
\cmidrule(lr){2-3} \cmidrule(lr){4-5}\cmidrule(lr){6-7}\cmidrule(lr){8-9}
$A\parallel Z$ & \rule{0pt}{3ex}$b^2\sin^{4}(\phi)+c^2\cos^{4}(\phi)$ & $\tikzcenter{\input{eq/a_para}}$ & 0 &  & 0 &  & $f^2\sin^{2}(2\phi)$ & $\tikzcenter{\input{eq/sin_2_2phi_}}$ \\
(100) & $\frac{1}{4}(b^2+c^2)\sin^{2}(2\phi)$ & $\tikzcenter{\input{eq/a_perp}}$ & 0 &  & 0 &  & $f^2\cos^{2}(2\phi)$ & $\tikzcenter{\input{eq/cos_2_2phi_}}$ \\
$B\parallel Z$ & \rule{0pt}{3ex} $c^2\sin^{4}(\phi)+a^2\cos^{4}(\phi)$ & $\tikzcenter{\input{eq/b_para}}$ & 0 &  & $e^2\sin^{2}(2\phi)$ & $\tikzcenter{\input{eq/sin_2_2phi_}}$ & 0 &  \\
(010) & $\frac{1}{4}(a^2+c^2)\sin^{2}(2\phi)$ & $\tikzcenter{\input{eq/b_perp}}$ & 0 &  & $e^2\cos^{2}(2\phi)$ & $\tikzcenter{\input{eq/cos_2_2phi_}}$ & 0 &  \\
$C\parallel Z$ & \rule{0pt}{3ex} $a^2\sin^{4}(\phi)+b^2\cos^{4}(\phi)$ & $\tikzcenter{\input{eq/c_para}}$ & $d^2\sin^{2}(2\phi)$ & $\tikzcenter{\input{eq/sin_2_2phi_}}$ & 0 &  & 0 &  \\
(001) & $\frac{1}{4}(a^2+b^2)\sin^{2}(2\phi)$ & $\tikzcenter{\input{eq/c_perp}}$ & $d^2\cos^{2}(2\phi)$ & $\tikzcenter{\input{eq/cos_2_2phi_}}$ & 0 &  & 0 &  \\
$N\parallel Z$ & \rule{0pt}{3ex} $b^2\cos^{4}(\phi)+(a'+c')^2\sin^{4}(\phi)$ & $\tikzcenter{\input{eq/101_para}}$ & $d'^2\sin^{2}(2\phi)$ & $\tikzcenter{\input{eq/0.5sin_2_2phi_}}$ & $e'^2\sin^4(\phi)$ & $\tikzcenter{\input{eq/sin_4_phi_}}$ & $f'^2\sin^{2}(2\phi)$ & $\tikzcenter{\input{eq/0.5sin_2_2phi_}}$ \\
(101) & $\frac{1}{4}\left(b^2+(a'+c')^2\right)\sin^{2}(2\phi)$ & $\tikzcenter{\input{eq/101_perp}}$ & $d'^2\cos^{2}(2\phi)$ & $\tikzcenter{\input{eq/0.5cos_2_2phi_}}$ & $\frac{1}{4}e'^2\sin^{2}(2\phi)$ & $\tikzcenter{\input{eq/0.25sin_2_2phi_}}$ & $f'^2\cos^{2}(2\phi)$ & $\tikzcenter{\input{eq/0.5cos_2_2phi_}}$ \\ \midrule
 & \multicolumn{8}{r}{$a'=a\cos^{2}(\theta)$, $c'=c\sin^{2}(\theta)$, $d'=d\cos(\theta)$}\\
 & \multicolumn{8}{r}{$e'=e\sin(2\theta)$, $f'=f\sin(\theta)$, $\theta=\arctan\left(\frac{c_0}{a_0}\right)$
 }\\
 \bottomrule \bottomrule
\end{tabular}
\label{tab:selection}
\end{table*}



\section{Methods}
\subsection{Raman spectroscopy}
Raman measurements were performed with a Horiba LabRAM HR Evolution Raman spectrometer using linearly polarized \SI{632.8}{\nano\metre} HeNe laser excitation in back-scattering geometry. A half-wave plate (HWP) on a motorized rotation stage allows switching of the polarization angle between $\hat{e}_{\rm i}$ and $\hat{e}_{\rm s}$ from 0° to 90°, enabling parallel and crossed polarization configurations. After reflection from a notch filter, the beam passes through a second HWP, which gradually rotates the polarization relative to the crystal axes with a step size of 10°. A 10$\times$ objective (Olympus, NA = 0.25) focuses the laser onto the sample and collects the back-scattered signal. Passing the scattered light also through the second HWP effectively simulates a rotation of the sample around the axis which is aligned with the $Z$ direction without physically moving it, thereby maintaining the probed spot. A fixed analyzer then selects the desired polarization before the scattered light is led through a \SI{200}{\micro\metre} wide pinhole and dispersed by a \SI{1800}{lines/mm} grating offering a spectral resolution of \SI{1.2}{\per\centi\metre}. Finally, the light is detected with a charge-coupled device (CCD).\\

\subsection{Sample characteristics}
Samples for the present study were prepared from bulk LaInO$_3$ single crystals grown from the melt by the Vertical Gradient Freeze (VGF) method at Leibniz-Institut für Kristallzüchtung \cite{galazka_melt_2021}, cut into wafers with the four main orientations (100), (010), (001) and (101) and polished afterwards. Ellipsometric measurements yield refractive indices of $n_{A}=2.0992$, $n_B=2.0667$ and $n_C=2.1330$ at an excitation wavelength of \SI{632.8}{\nano\metre}. More details on the growth process and structural quality of the samples can be found in Ref.\ \cite{galazka_melt_2021}.\\
In principle, backscattering from the (101) surface allows the observation of all Raman-active phonon modes, whereas pairs of \Ag{} and $\rm{B_{xg}}$ modes can be detected for the other orientations (cf.\ Tab.\ \ref{tab:selection}), enabling us to distinguish between modes with similar frequencies but different symmetries. Contrary to LIO grown on a BSO substrate, which shows the formation of differently oriented domains \cite{zupancic_role_2020}, the samples in this study are single-domain crystals, meaning the derived selection rules in Tab.\ \ref{tab:selection} apply and no mixed phonon signatures should be expected.

\subsection{DFT calculations}
\label{sec:DFT}
DFT as implemented in the \textit{Quantum Espresso} suite \cite{giannozzi_quantum_2009}, was used with Generalized Gradient Approximation (GGA) using Perdew-Burke-Ernzerhof (PBE) functionals to calculate the phonon dispersion and derive the oscillation patterns of the phonon modes at the $\Gamma$ point. Pseudopotentials from the \textit{Materials Cloud Precision} library, SSSP PBE Precision v1.3.0 ~\cite{prandini2018precision}, considering 5d$^1$6s$^2$ (La), 4d$^{10}$5s$^{2}$5p$^1$ (In), and 2s$^2$2p$^4$ (O) as valence configuration, were used with a kinetic cutoff-energy of \SI{80}{Ry} (\SI{800}{Ry}) for the wave function (charge density and potential). 
Electronic and vibrational properties were calculated on a regular $6\times4\times6$ $k$- and $q$-point grid. The optimized lattice parameters, $a_0 = \SI{6.015}{\angstrom}$, $b_0 = \SI{8.348}{\angstrom}$, and $c_0 = \SI{5.776}{\angstrom}$, are found to be slightly larger than experimental XRD values~\cite{galazka_melt_2021, hartley_experimental_2021}, consistent with the known \textit{underbinding} of GGA-PBE–based calculations~\cite{Skelton2015}. The resulting underestimation of bond strengths leads to systematically reduced phonon energies; accordingly, the presented phonon dispersion was uniformly scaled by a factor of 1.023, as determined from comparison with experiment. 
The phonon density of states (pDOS) was computed on a $50\times50\times50$ $q$-point grid.


\section{Results and Discussion}

\begin{figure}[t]
\includegraphics*[width=\linewidth]{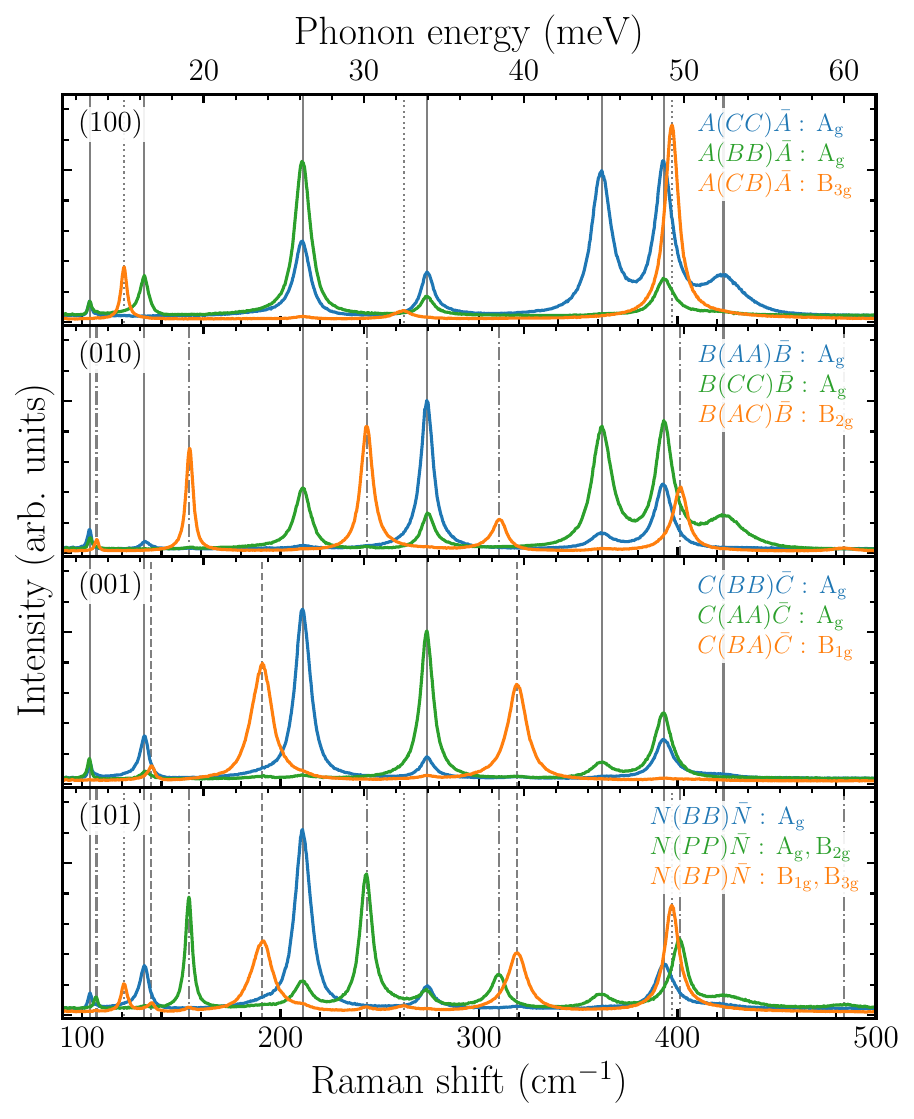}
\caption{Polarization-dependent Raman spectra of LIO, recorded with incident and scattered light directed along the $A$, $B$, $C$, and $N$ crystal axes (from top to bottom). Blue and green spectra correspond to parallel polarization configurations, with the polarization vectors rotated by 90° within the $XY$ plane, while orange spectra represent the crossed polarization configuration. The axis $P$ denotes the [$\bar{1}01$] crystal direction. Vertical solid (\Ag{}), dashed (\Beg), dash-dotted (\Bzg{}), and dotted (\Bdg{}) lines mark the positions of Raman-active phonon modes.}
\label{fig:Raman}
\end{figure}

Figure ~\ref{fig:Raman} presents the polarized Raman spectra obtained by backscattering from the (100), (010), (001) and (101) crystal faces in three different configurations. They are expressed in Porto's notation $k_{\rm{i}}(e_{\rm{i}}\,e_{\rm{s}})k_{\rm{s}}$, where $k_{\rm{i,s}}$ are the vectors of the incident and scattered light. The polarization vectors $e_{\rm{i,s}}$ are aligned with relevant crystallographic axes, ensuring that each mode is either suppressed or at its local maximum intensity with respect to the polarization angle. Accordingly, the rotational angle $\phi$ to be considered in Tab.\ \ref{tab:selection} is 0° (blue) or 90° (green) for parallel configuration, and 0° (orange) for crossed configuration. Thus, the symmetry of the observed phonon peaks can be directly determined on the basis of the selections rules given in Tab \ref{tab:selection}. Specifically, \Ag{} modes appear on all faces in parallel but never in crossed configuration, with their intensities depending on the tensor elements $a$, $b$ and $c$. In contrast, \Beg{} and \Bdg{} modes are exclusively visible in crossed configuration. \Beg{} modes appear on the (001) and (101) faces, whereas \Bdg{} modes are found on the (100) and (101) faces. On the (101) face, the corresponding intensity is reduced by approximately a factor of two compared to the (001) and (100) faces, respectively. \Bzg{} modes are visible only on the (010) face in crossed configuration and on the (101) face in parallel-90° configuration, with nearly equal intensities in both cases. Only peaks that satisfy these conditions are considered as first-order Raman active modes. Additional weak peaks are disregarded, which arise from a slight misalignment of up to $\pm3.3$° between the crystal and lab coordinate system in the $XY$ plane, from the finite efficiencies of the polarizers and HWPs, or from the nonzero NA of the used objective.
By careful analysis of the spectra in Fig.\ \ref{fig:Raman}, one can identify 19 of the 24 active Raman modes, of which many overlap and are indistinguishable in unpolarized measurements. The positions of all observed modes are indicated by solid, dashed, dash-dotted and dotted vertical lines for \Ag{}, \Beg, \Bzg{} and \Bdg{} symmetry, respectively, and are summarized in Tab.\ \ref{tab:results}. The remaining five modes are expected to be either obscured by a mode of the same symmetry or very weak in intensity and are discussed later with the introduction of DFT results.

\begin{figure}[t]
\includegraphics*[width=\linewidth]{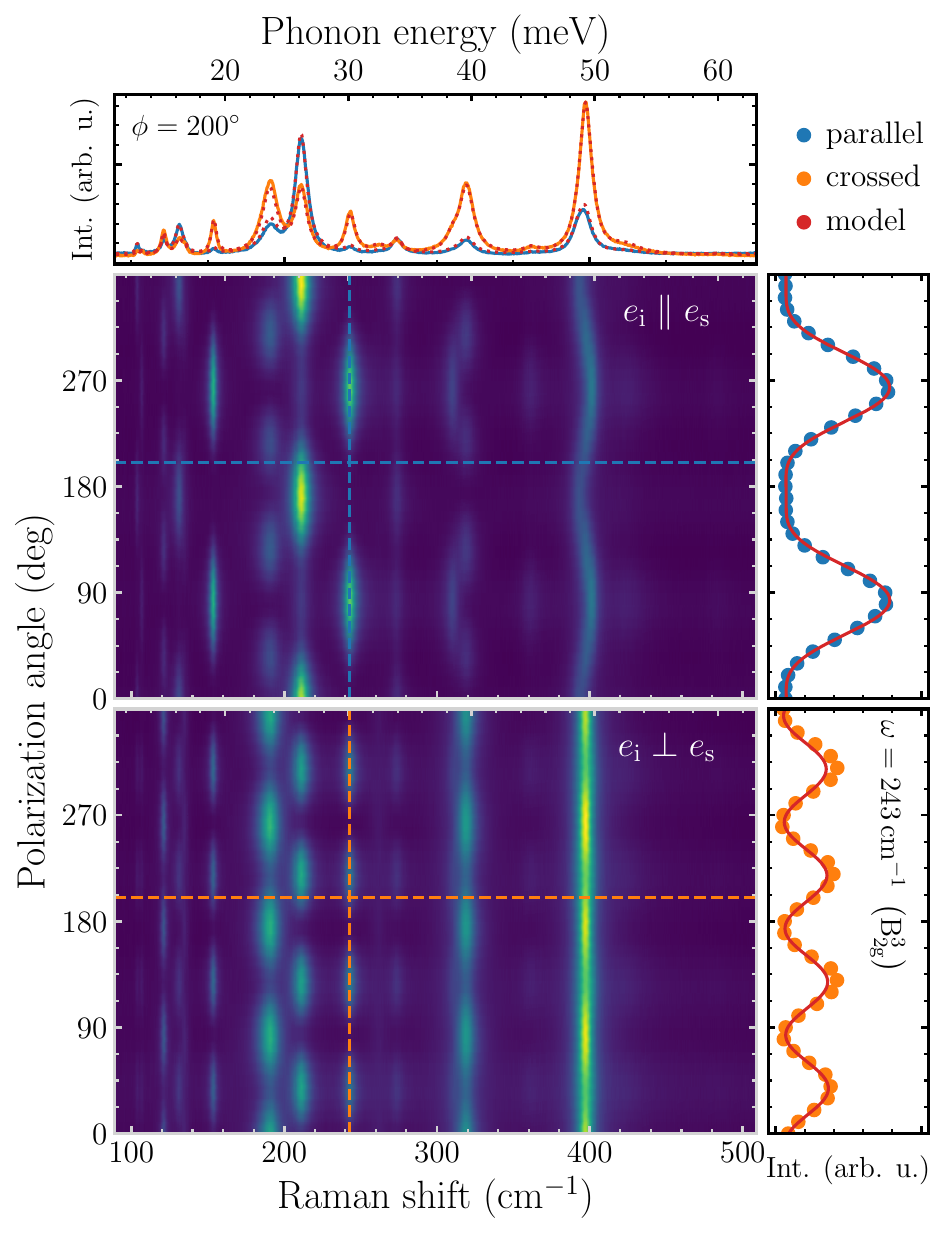}
\caption{Polarization-angle resolved Raman spectra of the (101) plane. The upper (lower) contour plot as well as blue (orange) lines and symbols correspond to measurements in parallel (crossed) polarization configuration. Red curves represent results from the hyperspectral fitting procedure. The panels above and to the right display slices of the data at 200° and \SI{243}{\per\centi\metre} ($\rm{B^3_{2g}}$ mode), respectively.}
\label{fig:contour}
\end{figure}

Full scans of polarization-angle resolved measurements on the (101)-plane are shown in Fig.\ \ref{fig:contour} and in the supplementary information for the other faces. For reference, the top panel displays the spectrum at $\phi=200$° where most of the modes are visible (analogous to Fig.\ \ref{fig:Raman}), while intensities at a fixed wavenumber ($\omega = \SI{243}{\per\centi\metre}$, $\rm{B^3_{2g}}$ mode) are plotted on the right for both parallel and crossed configurations. A rotational misalignment between the lab coordinate system and the crystal axes can be inferred from the systematic offsets of all peak maxima relative to their expected positions calculated from Tab.\ \ref{tab:selection}, and was later quantified from the fit as $-6.7$°. Similarly, this offset is $-20.9$° (100), 11.7° (010) and $-6.9$° (001) for the other faces in the SI. In addition, the background is modeled as $\cos^2(2\phi)$ and $\sin^2(2\phi)$ for parallel and crossed polarization configuration, respectively, as it shows a faint angle dependent intensity. We consider this to be an artefact of the experimental setup, but the exact mechanism remains unclear.
The red curves in Fig.\ \ref{fig:contour} represent slices from the modeling results that were determined by the procedure described in the following paragraph.\\
To obtain the Raman tensor elements from polarization dependent Raman data, the common approach is to analyze each spectrum separately for a fixed geometry and polarization configuration at different rotation angles. Each spectrum is fitted with a set of line shapes, here Lorentz functions, each corresponding to a specific phonon mode. From the fitting results, the amplitudes, i.\,e., integrated peak areas, are extracted and fitted with the intensity profiles given by Eq.\ (\ref{eq:intensity_biref}). However, this method becomes unreliable when phonon modes are close in frequency, such as the $\rm{A^2_g}$ and $\rm{B^1_{1g}}$ ($\approx \SI{130}{\per\centi\metre}$) or $\rm{A^6_g}$, $\rm{B^4_{3g}}$ and $\rm{B^5_{2g}}$ ($\approx \SI{400}{\per\centi\metre}$) modes, due to the strong correlation of the peak-describing parameters.
To overcome these limitations, a multidimensional (hyperspectral) fitting procedure was employed, that incorporates every spectrum measured across every geometry, in both parallel and crossed polarization, and at all rotation angles in one single global fit. In this approach, the angle-dependent intensity profiles from Tab.\ \ref{tab:selection} serve as modulation functions for each phonon line shape. This procedure constrains the evolution of the amplitude parameter and helps to separate overlapping modes by also exploiting information from multiple geometries and polarization configurations. More information about this method can be found in Ref.\ \cite{beta_GaO}.
The signs of the Raman tensor elements $b$, $d$, $e$ and $f$ cannot be determined as they appear only in squared form in the selection rules. In addition, birefringence allows the signs of $a$ and $c$ to be fixed only relative to one another, meaning they are either identical or opposite. To remove this ambiguity, $a$ is arbitrarily set to be positive. The fitted phonon frequencies and Raman tensor elements, divided by the square root of the Raman shift-dependent prefactor $F(\omega_{\rm{P}})$, are listed in Tab.\ \ref{tab:results}.

\begin{figure*}[th]
\includegraphics*[width=\linewidth]{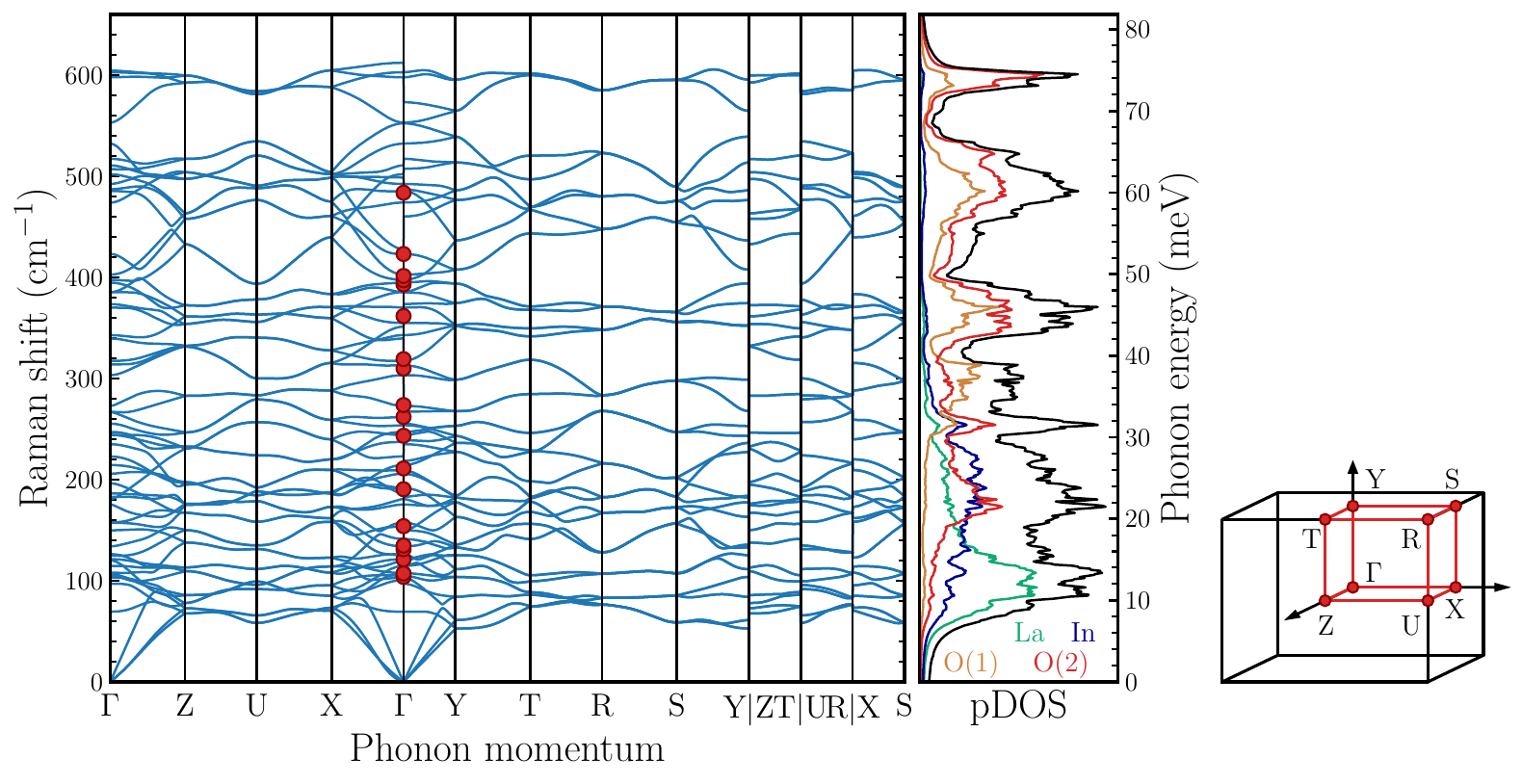}
\caption{Phonon dispersion of LIO along the high symmetry path displayed in the first Brillouin zone on the right side. Lines indicate DFT calculations and symbols mark experimentally obtained results. The right panel displays the site-resolved partial phonon density of states (pDOS) together with the total pDOS plotted in black.}
\label{Tor:fig:DFT}
\end{figure*}

\begin{table*}[t]
\caption{Frequencies of the Raman active $\Gamma$-point phonons in \si{\per\centi\metre} as obtained from the multidimensional fit, compared to results from DFT, as well as relative Raman tensor elements divided by the square root of the Raman shift-dependent intensity prefactor $F(\omega_{\rm{P}})$ from Eq. (\ref{eq:prefactor}) and normalized to 1000.}
\begin{tabular}{ccccc<{\l}>{\l}ccc<{\l}>{\l}ccc<{\l}>{\l}ccc}
\toprule \toprule
\multicolumn{5}{c}{\Ag} & \multicolumn{3}{c}{\Beg} & \multicolumn{3}{c}{\Bzg} & \multicolumn{3}{c}{\Bdg} \\
Exp.\ & DFT & $a$ & $\mathopen|b\mathopen|$ & $c$ & Exp.\ & DFT & $\mathopen|d\mathclose|$ & Exp.\ & DFT & $\mathopen|e\mathclose|$ & Exp.\ & DFT & $\mathopen|f\mathclose|$ \\ 
\cmidrule(lr){1-5}\cmidrule(lr){6-8}\cmidrule(lr){9-11}\cmidrule(lr){12-14}
104 & 103 & 860 & 693 & $-743$ & 135 & 120 & 397 & 107 & 105 & 568 & 121 & 114 & 953\\
131 & 126 & 320 & 769 & 4.3 & 191 & 180 & 603 & 154 & 157 & 1000 &  262  & 244 &  52.8\\
211 & 206 & 3.5 & 703 & $-499$ & 319 & 317 & 195 & 243 & 247 & 513 & -- & 385 & -- \\
274 & 273 & 470 & 161 & $-247$ &  --  & 402 &  --  & 310 & 303 & 174 & 397 & 397 & 174\\
362 & 374 & 104 & 26.6 & $-296$ &  --  & 553 &  --  & 401 & 394 & 163 &  --  & 602 &  -- \\
393 & 386 & 186 & 131 & $-264$ &  &  &  & 484 & 492 & 25.6 &  &  & \\
423 & 423 & 9.0 & 36.0 & 119 &  &  &  &  --  & 597 &  --  &  &  & 
 \\
\bottomrule \bottomrule
\label{tab:results}
\end{tabular}
\end{table*}

Figure \ref{Tor:fig:DFT} shows the DFT-calculated phonon dispersion along the high-symmetry path $\Gamma$ZUX$\Gamma$YTRSY$\vert$ZT$\vert$UR$\vert$XS. Additionally, the phonon density of states is depicted in the right panel of the figure. Owing to their masses, low-frequency modes are dominated by motions of the La atoms, followed by a broad distribution of In atoms. Modes with frequencies higher than $\approx\SI{300}{\per\centi\metre}$ consist almost entirely of O vibrations. The obtained $\Gamma$ point phonon energies, which are uniformly scaled by a factor of 1.023 as justified in Sec.\ \ref{sec:DFT}, align very well with the experiment (Tab.\ \ref{tab:results}).\\
The results from DFT further indicate that the final modes to be observed experimentally are the $\rm{B^4_{1g}}$, $\rm{B^5_{1g}}$, $\rm{B^7_{2g}}$, $\rm{B^3_{3g}}$ and $\rm{B^5_{3g}}$ modes. These modes are O-dominated and, according to the atomic displacement patterns (cf.\ SI), exhibit strong stretching-like character. In perovskites that do not exhibit Jahn–Teller distortions, i.\ e.\ symmetry breaking due to spatially degenerate electronic ground states, or at least have small deviations in their B-O bond lengths, such as CaMnO$_3$ or metallic Pnma compounds, similar stretching modes are known to display weak or vanishing Raman intensities \cite{martin-carron_raman_2002}. Since In$^{3+}$ in LIO has an electronically non-degenerate 4d$^{10}$ electron configuration and therefore does not support a Jahn-Teller distortion, this effect might explain the absence of these modes in our measurements.  

\begin{figure}[t]
\begin{overpic}[width=0.49\linewidth,
                trim=7cm 10cm 40cm 0cm, clip]
               {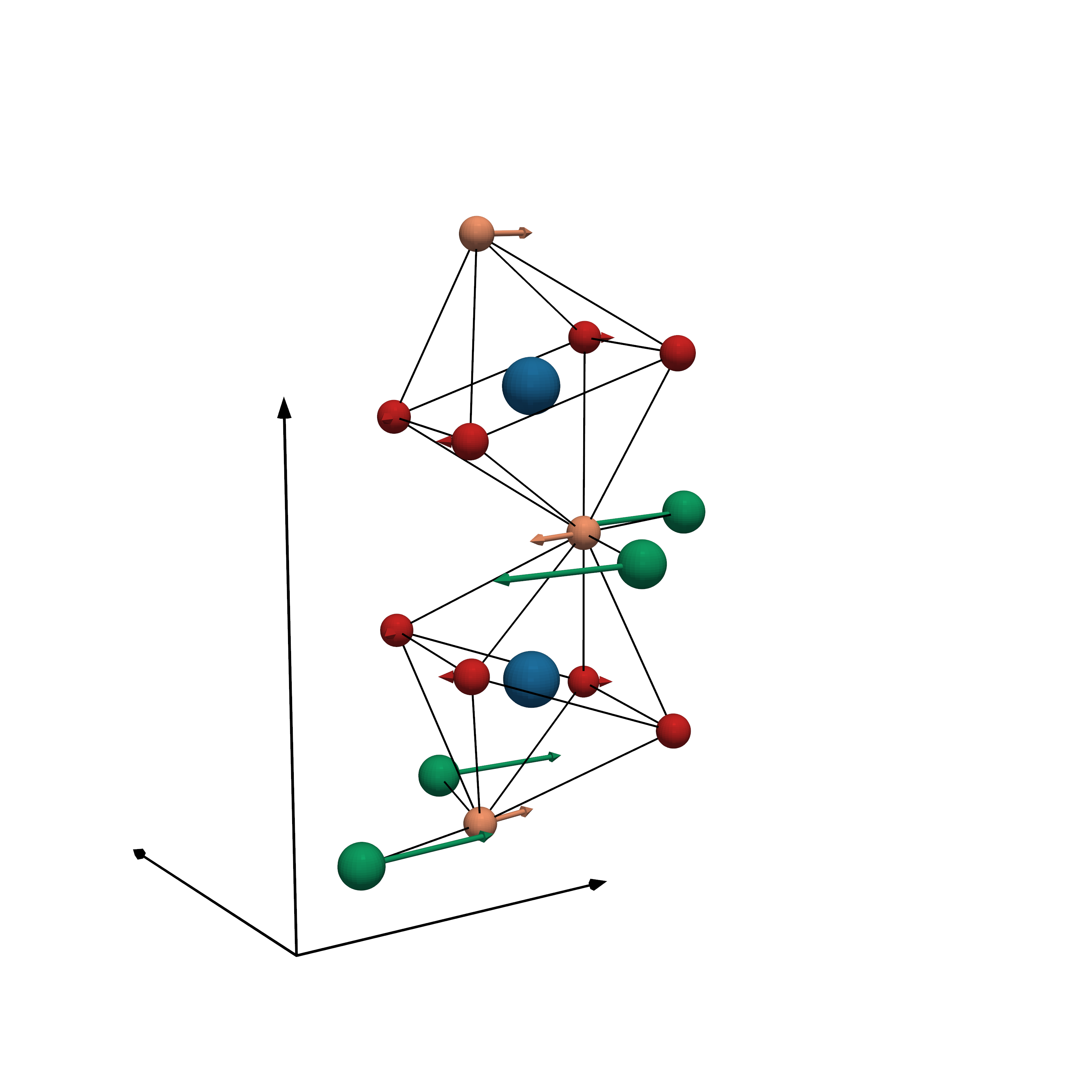}
    \put(43,84){\sffamily \Large $\rm{B^1_{2g}}$}
    \put(52,13){\sffamily $Z$}
    \put(5,18){\sffamily $X$}
    \put(22,64){\sffamily $Y$}
\end{overpic}
\hfill
\begin{overpic}[width=0.49\linewidth,
                trim=15cm 10cm 32cm 0cm, clip]
               {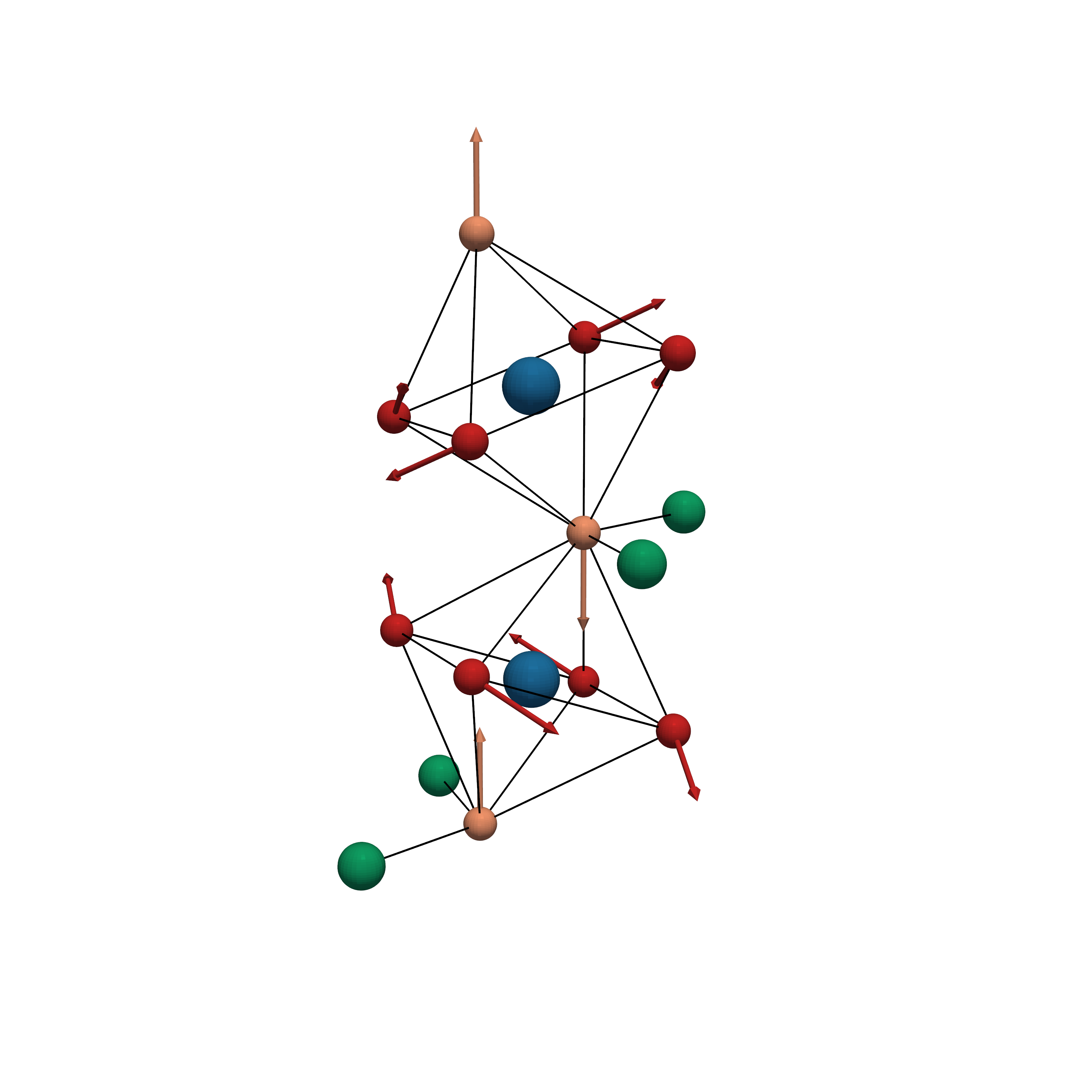}
    \put(35,84){\sffamily\Large $\rm{B^3_{1g}}$}
\end{overpic}

\caption{Phonon displacement patterns of the $\rm{B^1_{2g}}$ and $\rm{B^3_{1g}}$ modes in LIO. La-, In-, O(1) and O(2)-atoms are depicted in green, blue, orange and red, respectively.}
\label{Tor:fig:Eigenvec}
\end{figure}
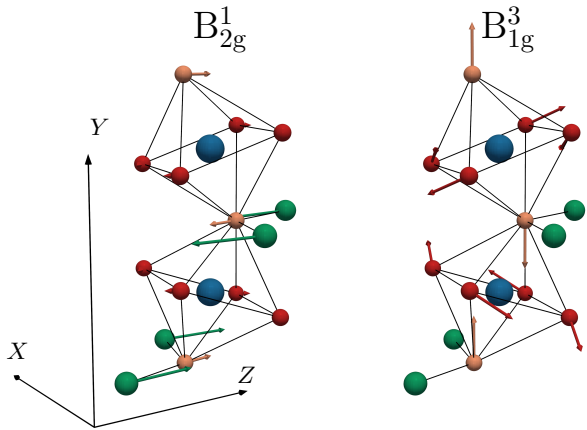

Figure \ref{Tor:fig:Eigenvec} shows the oscillation patterns of the $\rm{B^1_{2g}}$ and $\rm{B^3_{1g}}$ modes at the $\Gamma$ point, with the remaining modes provided in the SI. The $\rm{B^1_{2g}}$ mode is dominated by oscillations of the La atoms, whereas the $\rm{B^3_{1g}}$ mode exclusively involves O sites oscillations. Consequently, these two modes might serve as sensitive indicators to be analyzed for specific crystal modifications, such as alloying in La$_{(1-x)}$Ga$_x$InO$_3$ or doping on O sites.
Furthermore, the Pnma space group of ABO$_3$ perovskites dictates specific characteristics of the atomic motions: Following the site symmetry of 4c ($\rm{C_s = \{E,\, m_{010}\}}$), displacements of A- and O(1)-atoms are constrained to the $XZ$-plane for \Ag{} and \Bzg{} modes, while \Beg{} and \Bdg{} modes are confined along the $Y$-direction. In contrast, there are no restrictions for the oscillations of O(2)-atoms, as their site symmetry of 8d ($\rm{C_1 = \{E\}}$) only contains the unit operation. In many perovskites, such as LaMnO$_3$ or YMnO$_3$, however, the O(2) vibrations are nearly orientated either parallel or perpendicular to the B-O(2) bonds, which are, due to the shorter bonding lengths, significantly stronger than the A-O(2) bonds \cite{iliev_raman_1998}. In LIO, the difference between these two bond types is smaller, resulting in less correlated O(2)-displacements. Although some modes correspond to nearly pure vibrations of an isolated BO$_6$ octahedron, including breathing ($\rm{B^5_{3g}}$), tilting ($\rm{A^4_g}$), bending ($\rm{B^2_{3g}}$) and rotational ($\rm{B^2_{1g}}$) modes, many of the displacements patterns represent mixtures of these motions and thus cannot be uniquely labeled, unlike in similar Pnma perovskites.

\section{Summary}
Summarizing and concluding, we studied the lattice dynamics of orthorhombic LIO using polarization-angle resolved Raman spectroscopy in combination with DFT calculations. Measurements in backscattering geometry on multiple crystallographic surfaces allowed the identification and symmetry assignment of 19 of the 24 Raman-active $\Gamma$-point phonon modes. We discuss the remaining modes and relate their absence in our spectra to their strong stretching-like character. A multidimensional fitting approach was employed to analyze the angular dependence of the Raman scattering intensities and to extract the relative Raman tensor elements, even for closely spaced and overlapping modes.
First-principles DFT calculations were used to compute the phonon dispersion, phonon density of states, and atomic displacement patterns. The calculated $\Gamma$-point phonon frequencies were found to be in good overall agreement with the experiment.\\
These results provide a reference for future studies addressing strain, alloying, or defect-induced modifications of the lattice, as well as for investigations of phonon-related transport and optical processes in LIO-based oxide heterostructures and electronic devices.

\section{Acknowledgements}
Crystal growth of LaInO$_3$ was supported by the Leibniz Senatsausschuss Wettbewerb (SAW) project BASTET. The authors thank Andreas Fiedler (Leibniz-Institut für Kristallzüchtung) for critical reading of the paper.

\bibliographystyle{apsrev4-2}
\section*{References}
\bibliography{LIO}


\input{supplemental}

\end{document}

%% file: eq/a_para.tex
\begin{tikzpicture}[scale=\scaleVar]
    \draw[domain=0:360,samples=100,smooth,variable=\t,color=teal]
        plot ({\t}:{\bVar^2*(sin(\t))^4 + \cVar^2*cos(\t)^4});
\end{tikzpicture}

%% file: eq/sin_2_2phi_.tex
\begin{tikzpicture}[scale=\scaleVar]
    \draw[domain=0:360,samples=100,smooth,variable=\t,color=teal]
        plot ({\t}:{(sin(2*\t))^2});
\end{tikzpicture}

%% file: eq/a_perp.tex
\begin{tikzpicture}[scale=\scaleVar]
    \draw[domain=0:360,samples=100,smooth,variable=\t,color=orange]
        plot ({\t}:{0.25*(\bVar^2+\cVar^2)*sin(2*\t)^2});
\end{tikzpicture}

%% file: eq/cos_2_2phi_.tex
\begin{tikzpicture}[scale=\scaleVar]
    \draw[domain=0:360,samples=100,smooth,variable=\t,color=orange]
        plot ({\t}:{(cos(2*\t))^2});
\end{tikzpicture}

%% file: eq/b_para.tex
\begin{tikzpicture}[scale=\scaleVar]
    \draw[domain=0:360,samples=100,smooth,variable=\t,color=teal]
        plot ({\t}:{\cVar^2*(sin(\t))^4 + \aVar^2*cos(\t)^4});
\end{tikzpicture}

%% file: eq/b_perp.tex
\begin{tikzpicture}[scale=\scaleVar]
    \draw[domain=0:360,samples=100,smooth,variable=\t,color=orange]
        plot ({\t}:{0.25*(\aVar^2+\cVar^2)*sin(2*\t)^2});
\end{tikzpicture}

%% file: eq/c_para.tex
\begin{tikzpicture}[scale=\scaleVar]
    \draw[domain=0:360,samples=100,smooth,variable=\t,color=teal]
        plot ({\t}:{\aVar^2*(sin(\t))^4 + \bVar^2*cos(\t)^4});
\end{tikzpicture}

%% file: eq/c_perp.tex
\begin{tikzpicture}[scale=\scaleVar]
    \draw[domain=0:360,samples=100,smooth,variable=\t,color=orange]
        plot ({\t}:{0.25*(\aVar^2+\bVar^2)*sin(2*\t)^2});
\end{tikzpicture}

%% file: eq/101_para.tex
\begin{tikzpicture}[scale=\scaleVar]
    \draw[domain=0:360,samples=100,smooth,variable=\t,color=teal]
        plot ({\t}:{(\bVar^2*(cos(\t))^4 + (\aVar/2 + \cVar/2)^2*sin(\t)^4)});
\end{tikzpicture}

%% file: eq/0.5sin_2_2phi_.tex
\begin{tikzpicture}[scale=\scaleVar]
    \draw[domain=0:360,samples=100,smooth,variable=\t,color=teal]
        plot ({\t}:{0.5*(sin(2*\t))^2});
\end{tikzpicture}

%% file: eq/sin_4_phi_.tex
\begin{tikzpicture}[scale=\scaleVar]
    \draw[domain=0:360,samples=100,smooth,variable=\t,color=teal]
        plot ({\t}:{(sin(\t))^4});
\end{tikzpicture}

%% file: eq/101_perp.tex
\begin{tikzpicture}[scale=\scaleVar]
    \draw[domain=0:360,samples=100,smooth,variable=\t,color=orange]
        plot ({\t}:{0.25*(\bVar^2+(\aVar/2+\cVar/2)^2)*sin(2*\t)^2});
\end{tikzpicture}

%% file: eq/0.5cos_2_2phi_.tex
\begin{tikzpicture}[scale=\scaleVar]
    \draw[domain=0:360,samples=100,smooth,variable=\t,color=orange]
        plot ({\t}:{0.5*(cos(2*\t))^2});
\end{tikzpicture}

%% file: eq/0.25sin_2_2phi_.tex
\begin{tikzpicture}[scale=\scaleVar]
    \draw[domain=0:360,samples=100,smooth,variable=\t,color=orange]
        plot ({\t}:{0.25*(sin(2*\t))^2});
\end{tikzpicture}

%% file: supplemental.tex
\clearpage
\newgeometry{textwidth=17.5cm, textheight=24cm} 
\onecolumngrid 
\appendix

\renewcommand\thefigure{
S\arabic{figure}}    
\setcounter{figure}{0}  

{
\LARGE\noindent Supplementary information for: \\
\large Lattice dynamics and complete polarization analysis of Raman-active modes in LaInO$_3$
}

\begin{overpic}[width=0.24\linewidth,
                trim=25cm 10cm 30cm 5cm, clip]
               {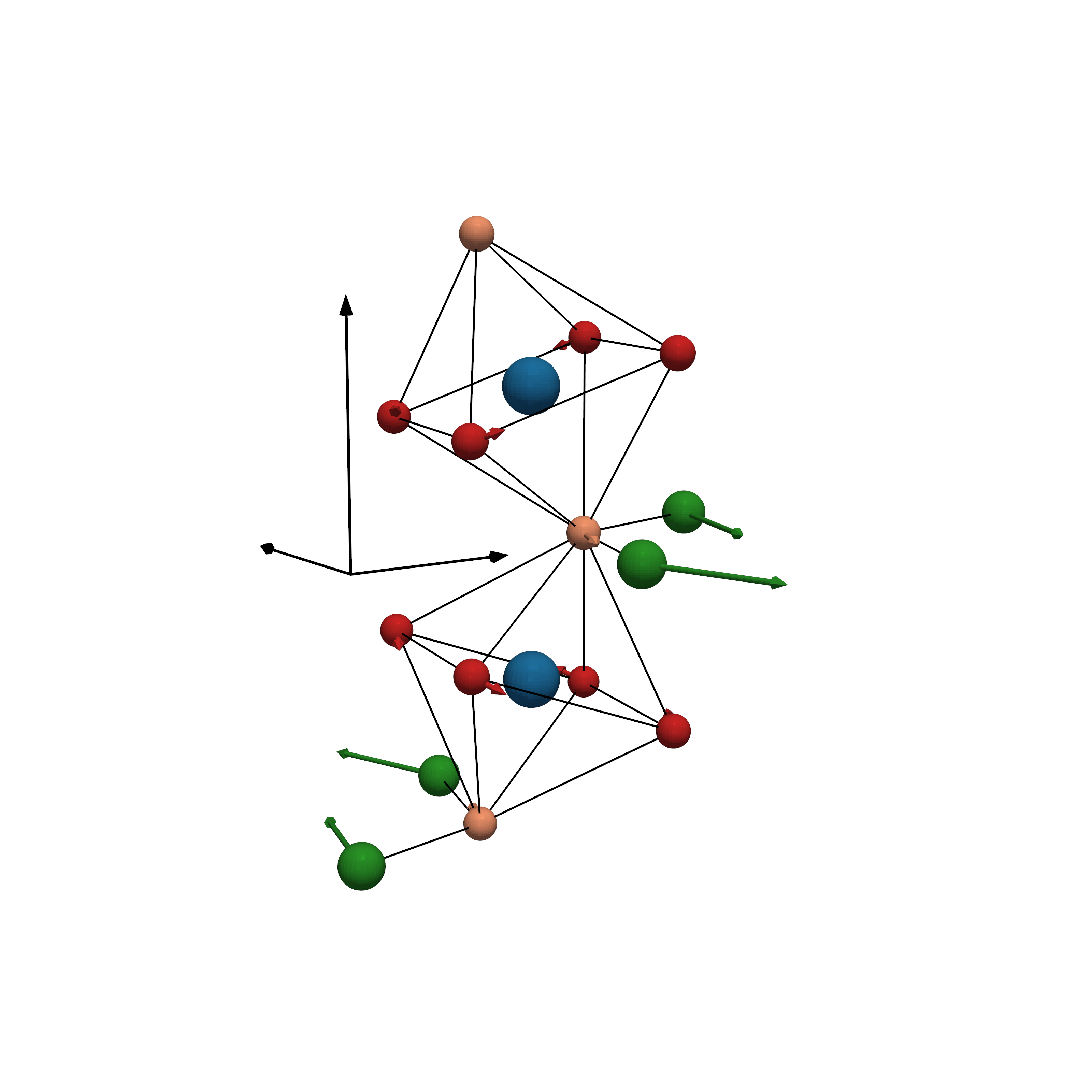}
    \put(28,84){\sffamily \Large $\rm{A^1_{g}}$}
    \put(21,49){\sffamily $Z$}
    \put(2,50){\sffamily $X$}
    \put(10,75){\sffamily $Y$}
\end{overpic}
\hfill
\begin{overpic}[width=0.24\linewidth,
                trim=25cm 10cm 30cm 5cm, clip]
               {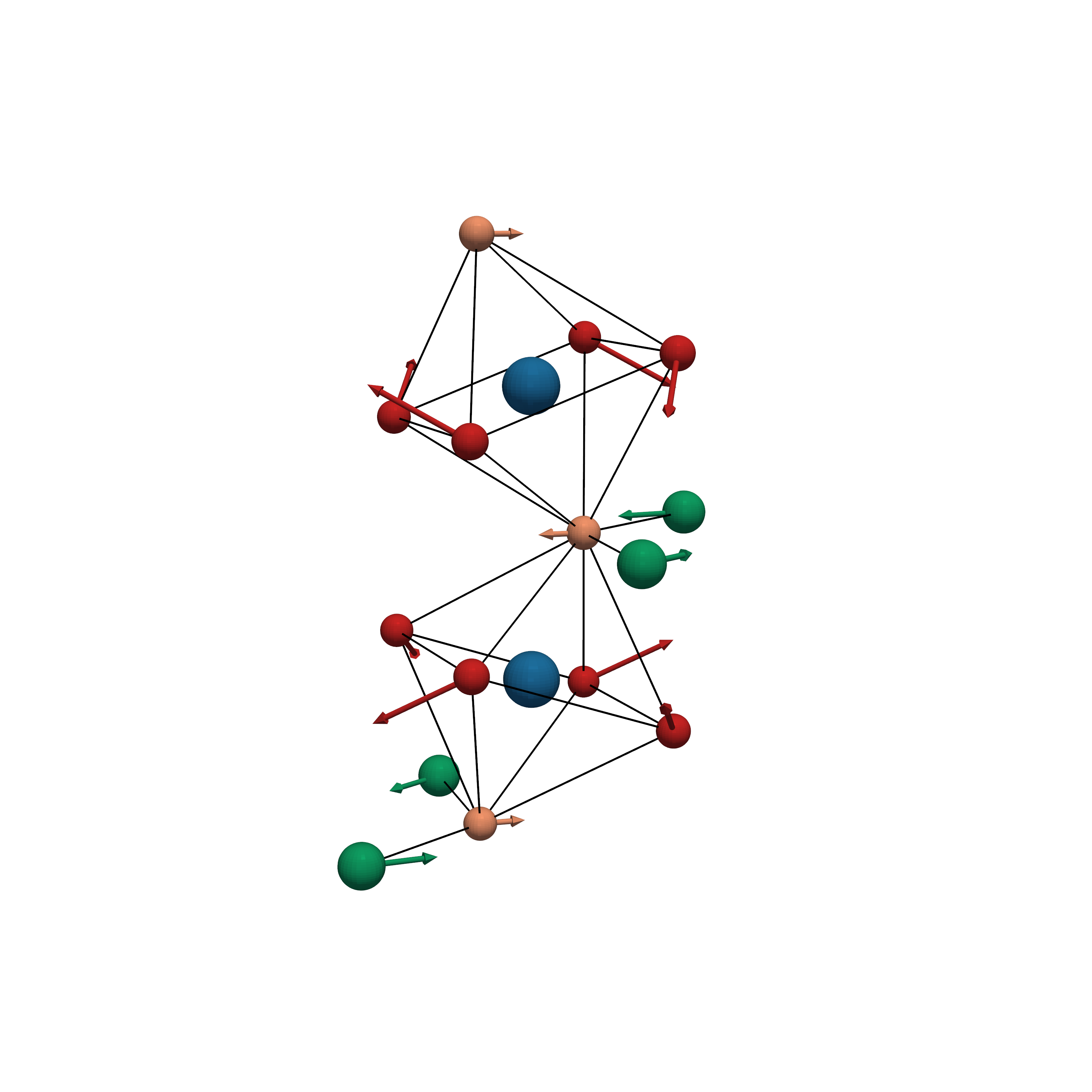}
    \put(28,84){\sffamily\Large $\rm{A^2_{g}}$}
\end{overpic}
\hfill
\begin{overpic}[width=0.24\linewidth,
                trim=25cm 10cm 30cm 5cm, clip]
               {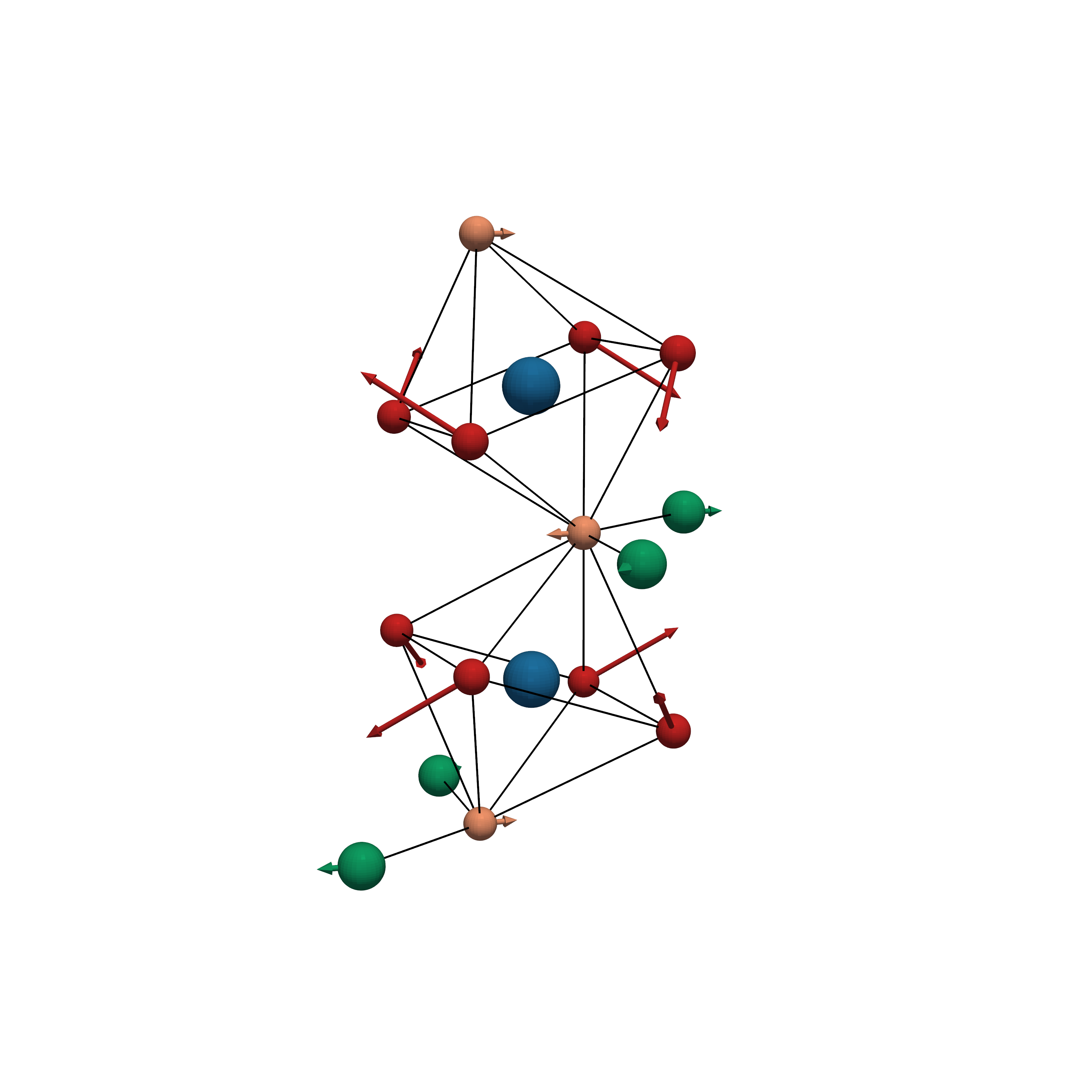}
    \put(28,84){\sffamily\Large $\rm{A^3_{g}}$}
\end{overpic}
\hfill
\begin{overpic}[width=0.24\linewidth,
                trim=25cm 10cm 30cm 5cm, clip]
               {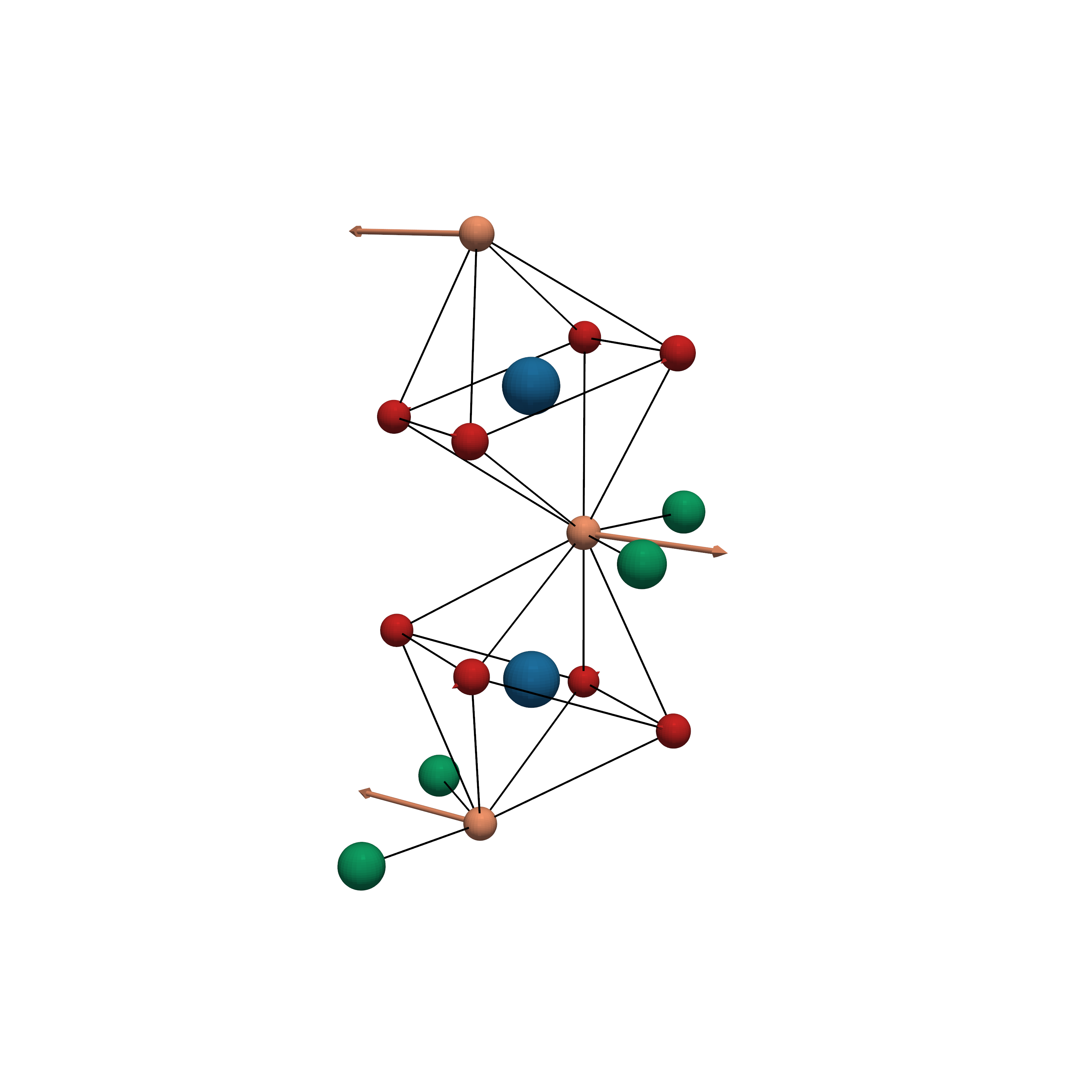}
    \put(28,84){\sffamily\Large $\rm{A^4_{g}}$}
\end{overpic}
\hfill \\

\begin{overpic}[width=0.24\linewidth,
                trim=25cm 10cm 30cm 5cm, clip]
               {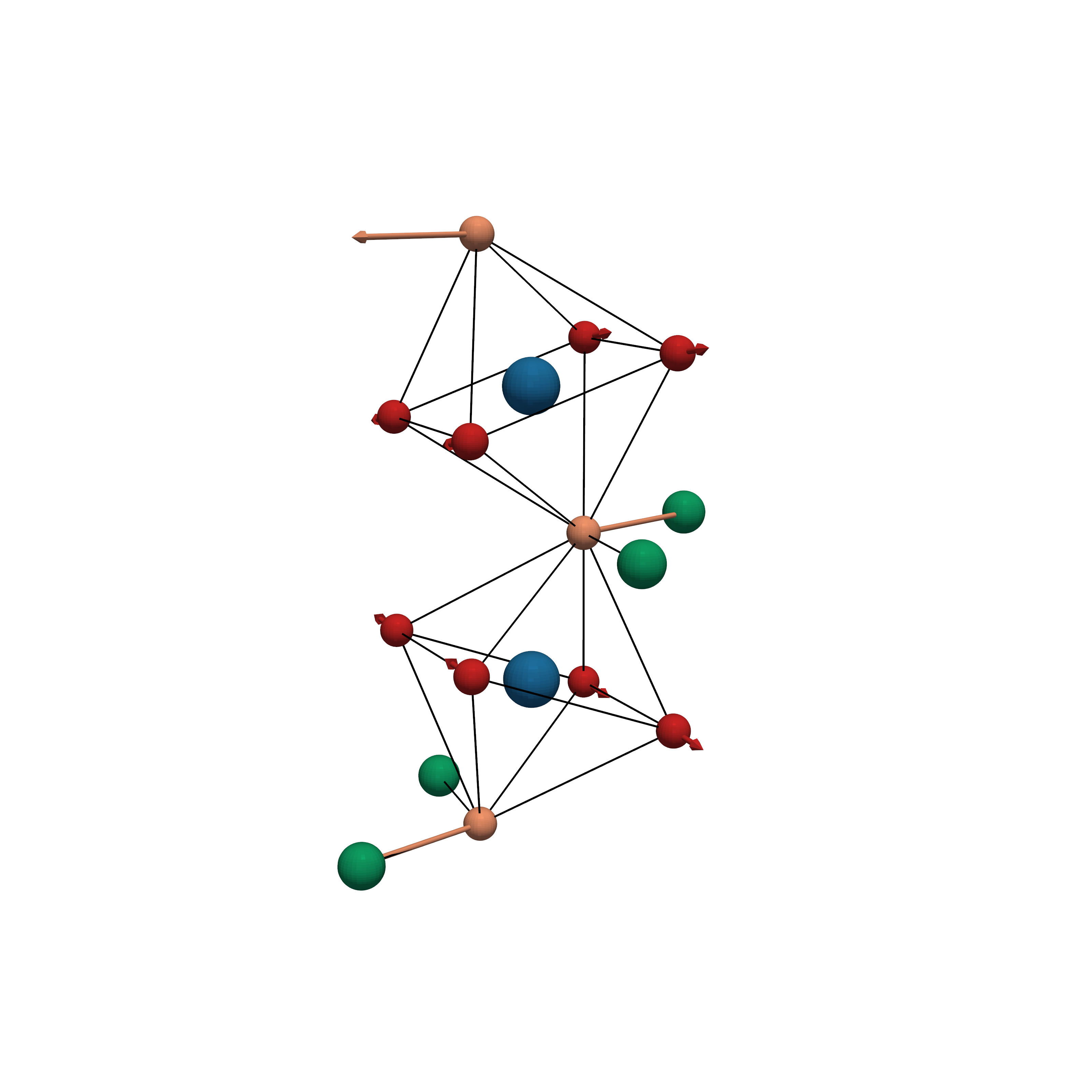}
    \put(28,84){\sffamily\Large $\rm{A^5_{g}}$}
\end{overpic}
\hfill
\begin{overpic}[width=0.24\linewidth,
                trim=25cm 10cm 30cm 5cm, clip]
               {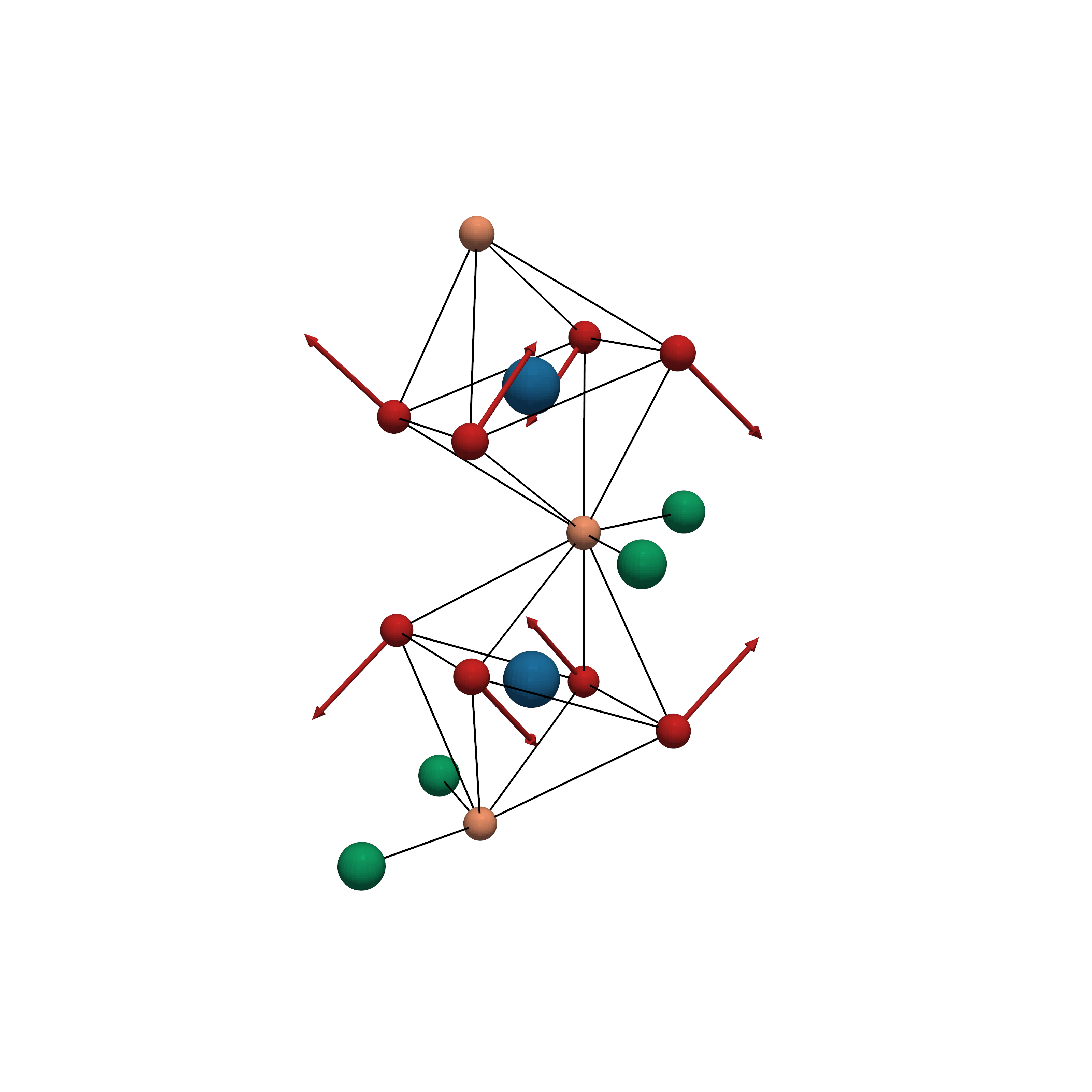}
    \put(28,84){\sffamily\Large $\rm{A^6_{g}}$}
\end{overpic}
\hfill
\begin{overpic}[width=0.24\linewidth,
                trim=25cm 10cm 30cm 5cm, clip]
               {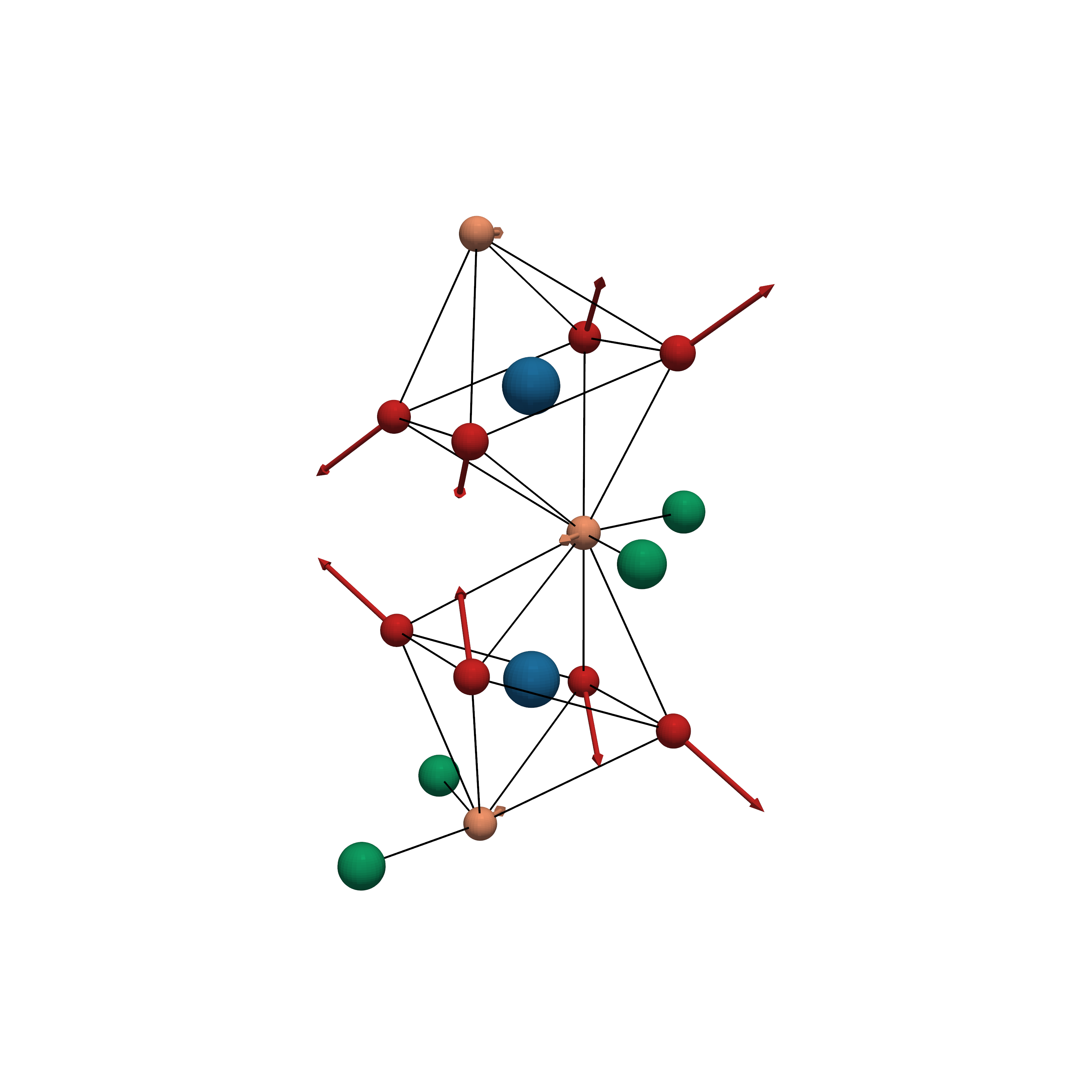}
    \put(28,84){\sffamily\Large $\rm{A^7_{g}}$}
\end{overpic}
\hfill
\begin{overpic}[width=0.238\linewidth,
                trim=25cm 10cm 30cm 5cm, clip]
               {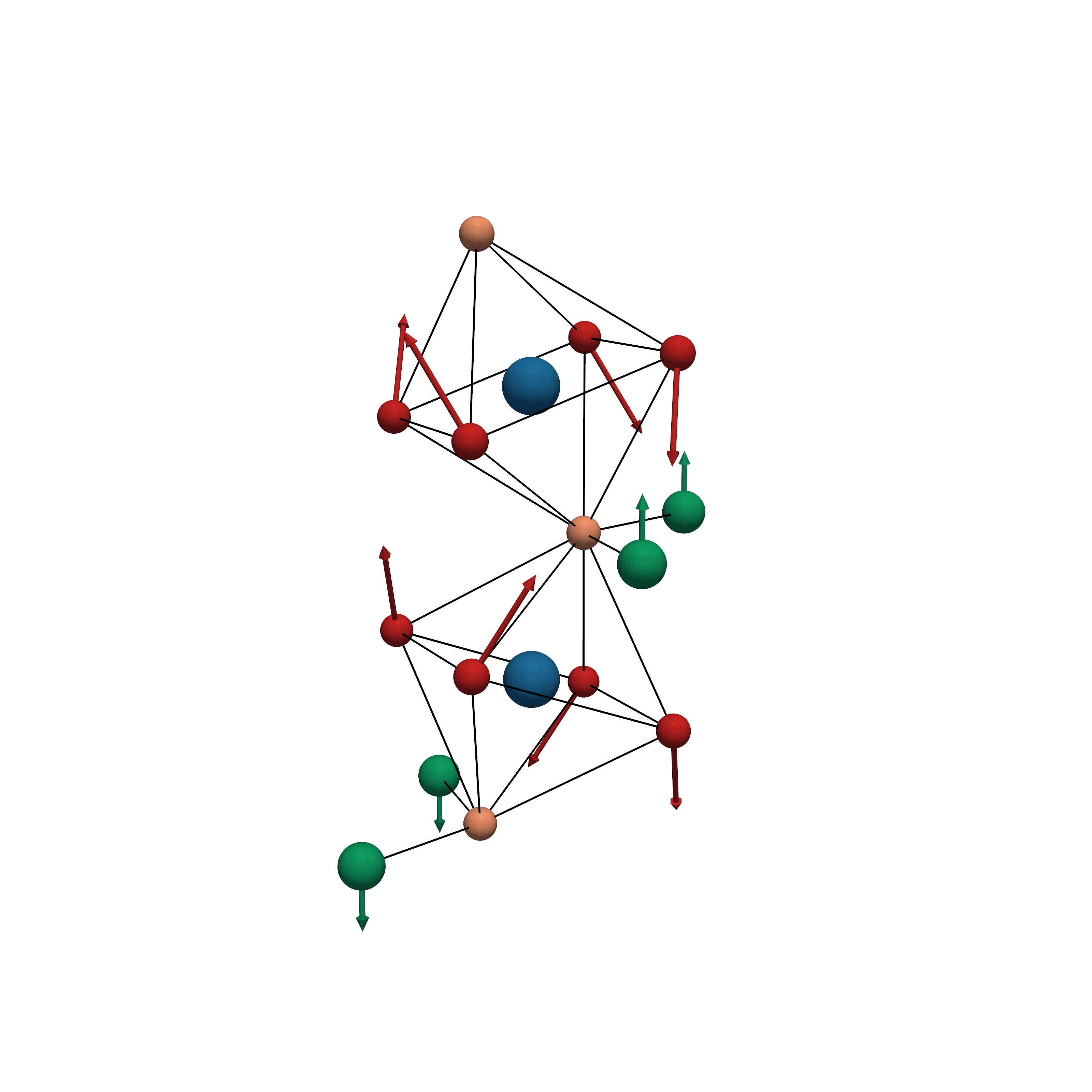}
    \put(28,84){\sffamily\Large $\rm{B^1_{1g}}$}
\end{overpic}
\hfill

\begin{overpic}[width=0.24\linewidth,
                trim=25cm 10cm 30cm 5cm, clip]
               {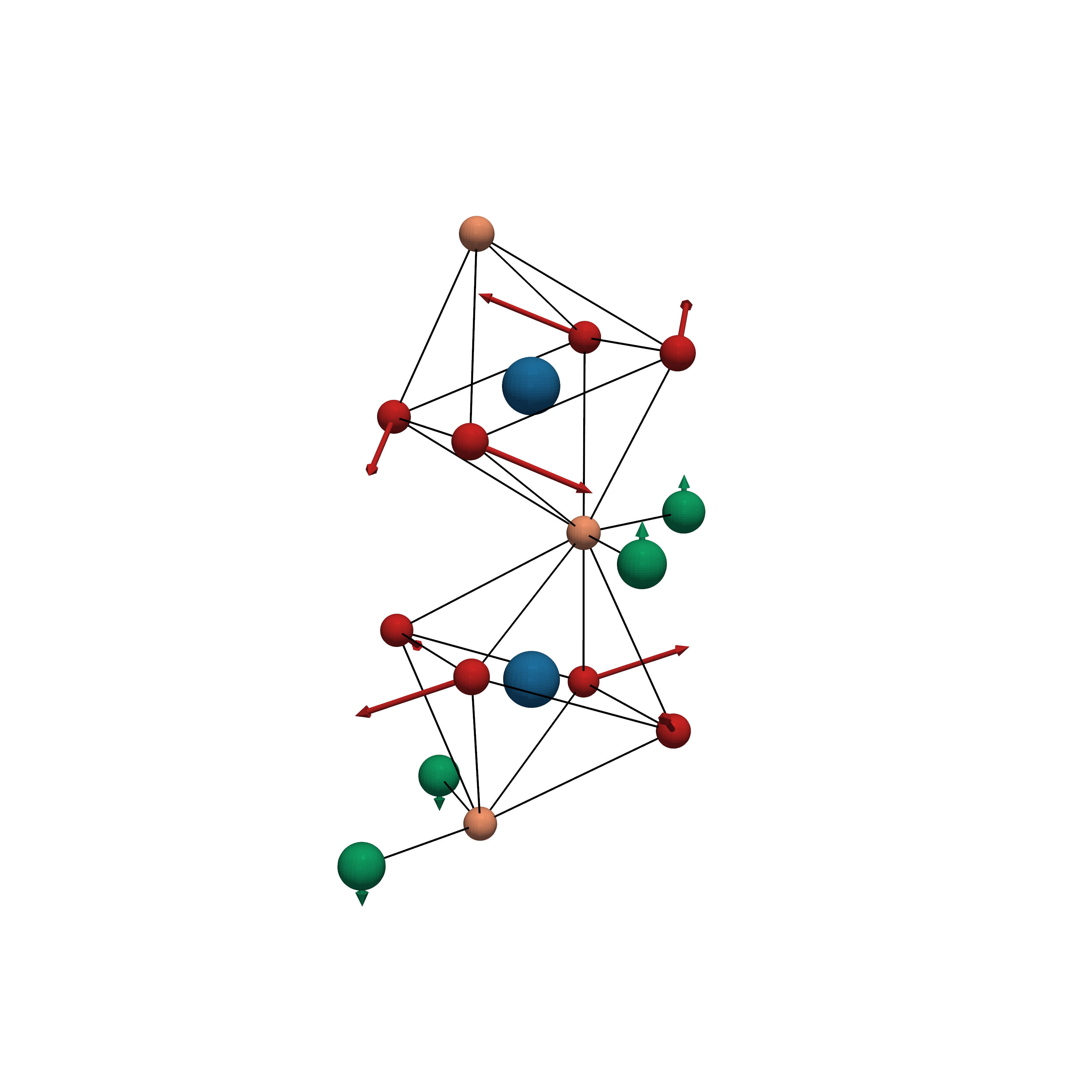}
    \put(28,84){\sffamily\Large $\rm{B^2_{1g}}$}
\end{overpic}
\hfill
\begin{overpic}[width=0.24\linewidth,
                trim=25cm 10cm 30cm 5cm, clip]
               {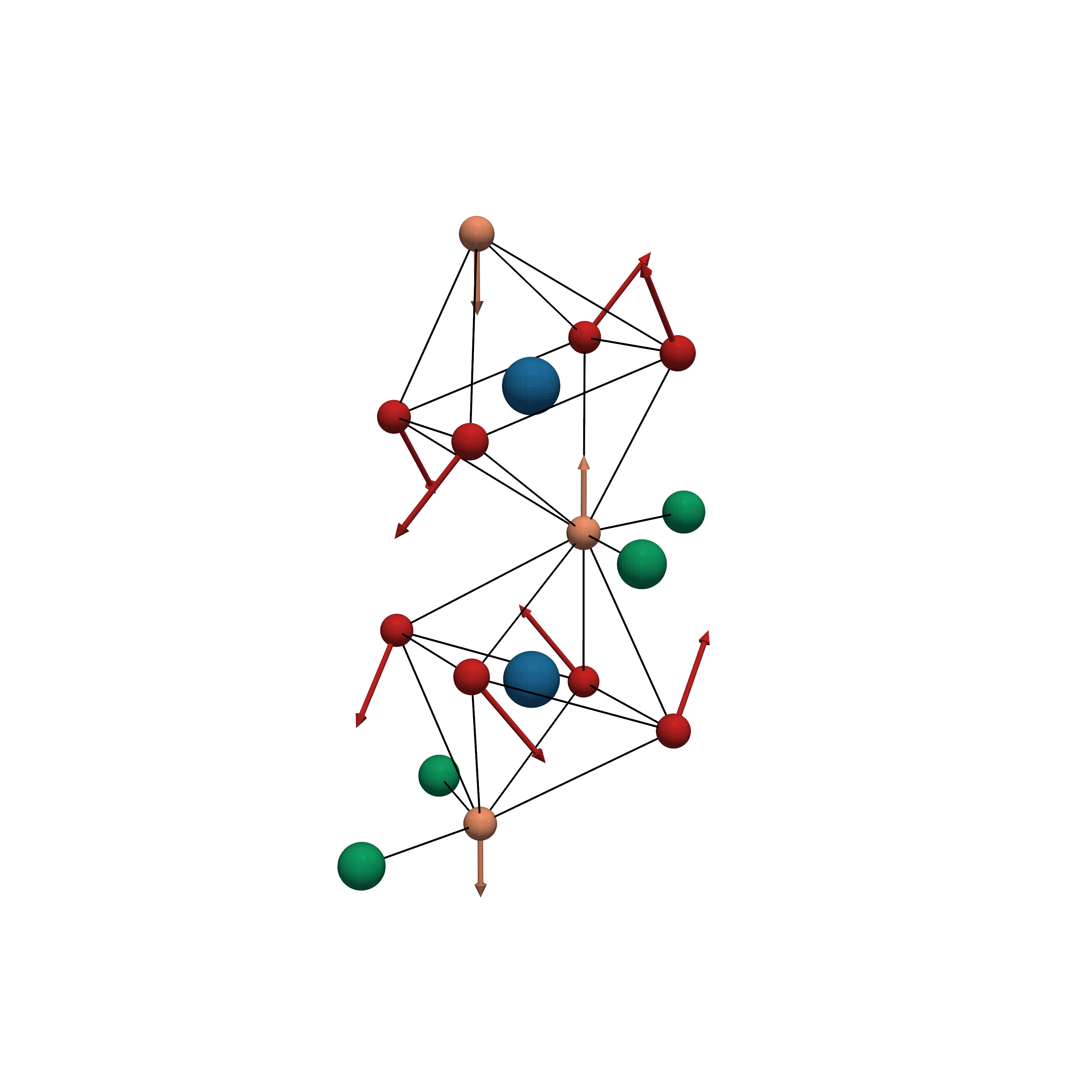}
    \put(28,84){\sffamily\Large $\rm{B^3_{1g}}$}
\end{overpic}
\hfill
\begin{overpic}[width=0.24\linewidth,
                trim=25cm 10cm 30cm 5cm, clip]
               {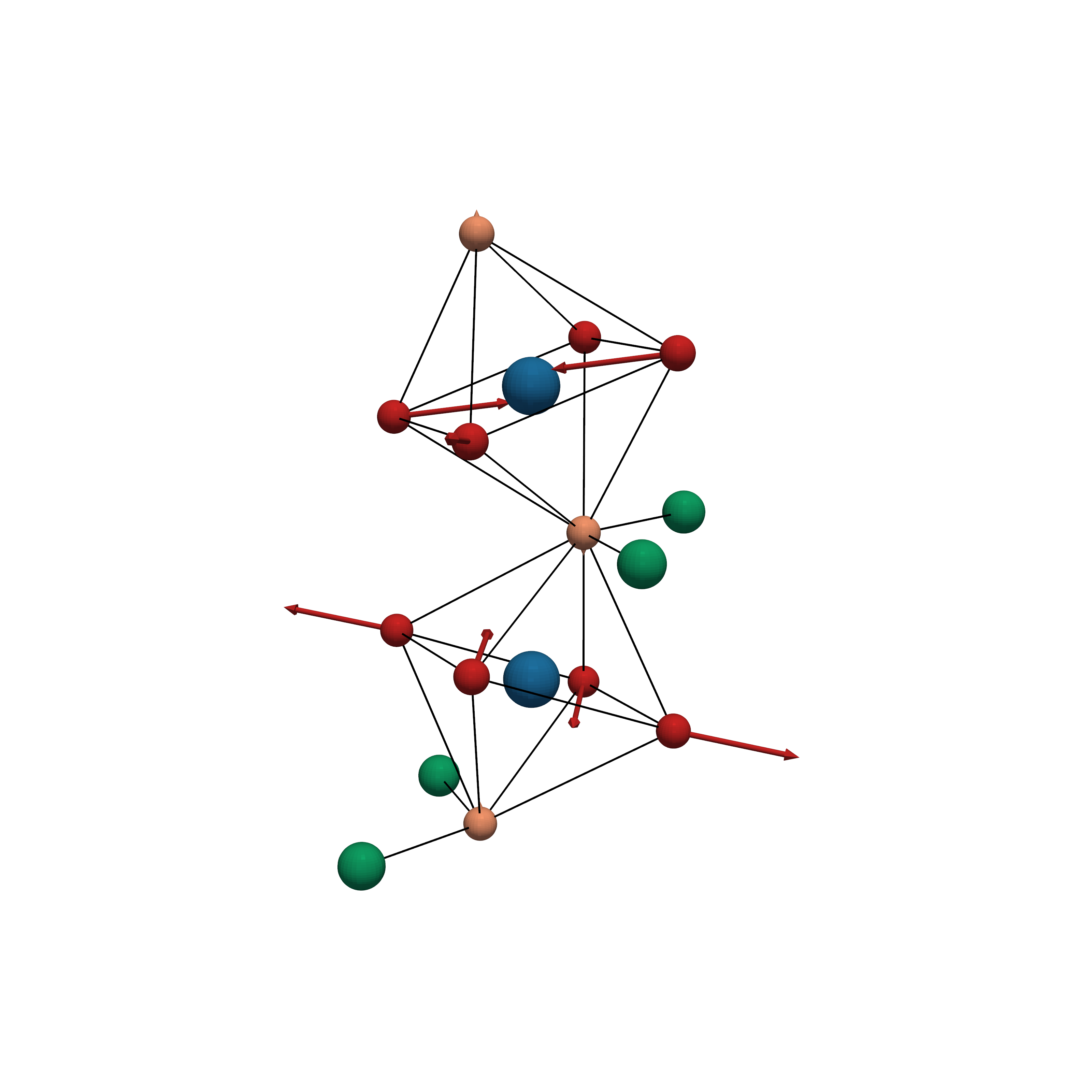}
    \put(28,84){\sffamily\Large $\rm{B^4_{1g}}$}
\end{overpic}
\hfill
\begin{overpic}[width=0.24\linewidth,
                trim=25cm 10cm 30cm 5cm, clip]
               {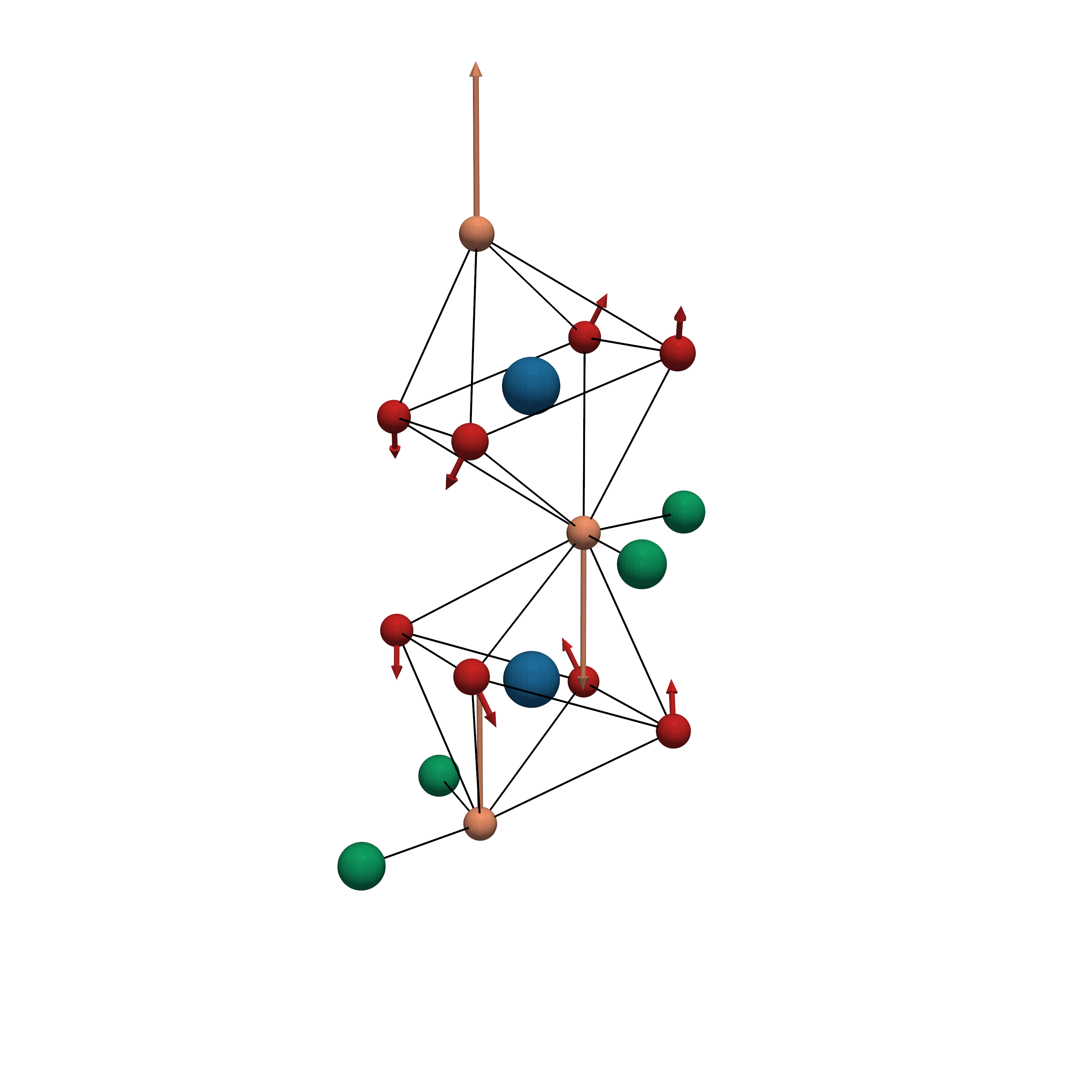}
    \put(28,84){\sffamily\Large $\rm{B^5_{1g}}$}
\end{overpic}
\hfill \\

\begin{overpic}[width=0.24\linewidth,
                trim=25cm 10cm 30cm 5cm, clip]
               {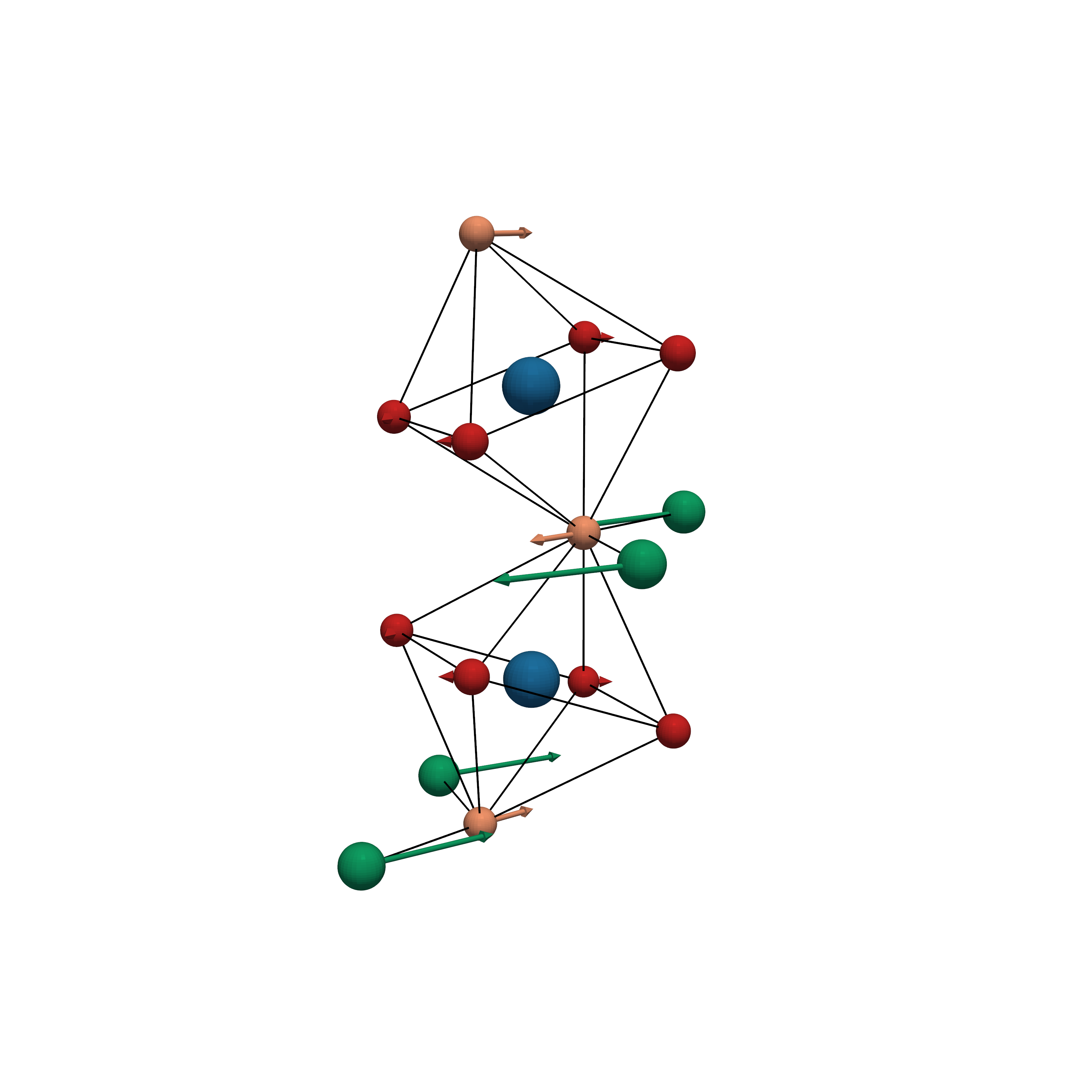}
    \put(28,84){\sffamily\Large $\rm{B^1_{2g}}$}
\end{overpic}
\hfill
\begin{overpic}[width=0.24\linewidth,
                trim=25cm 10cm 30cm 5cm, clip]
               {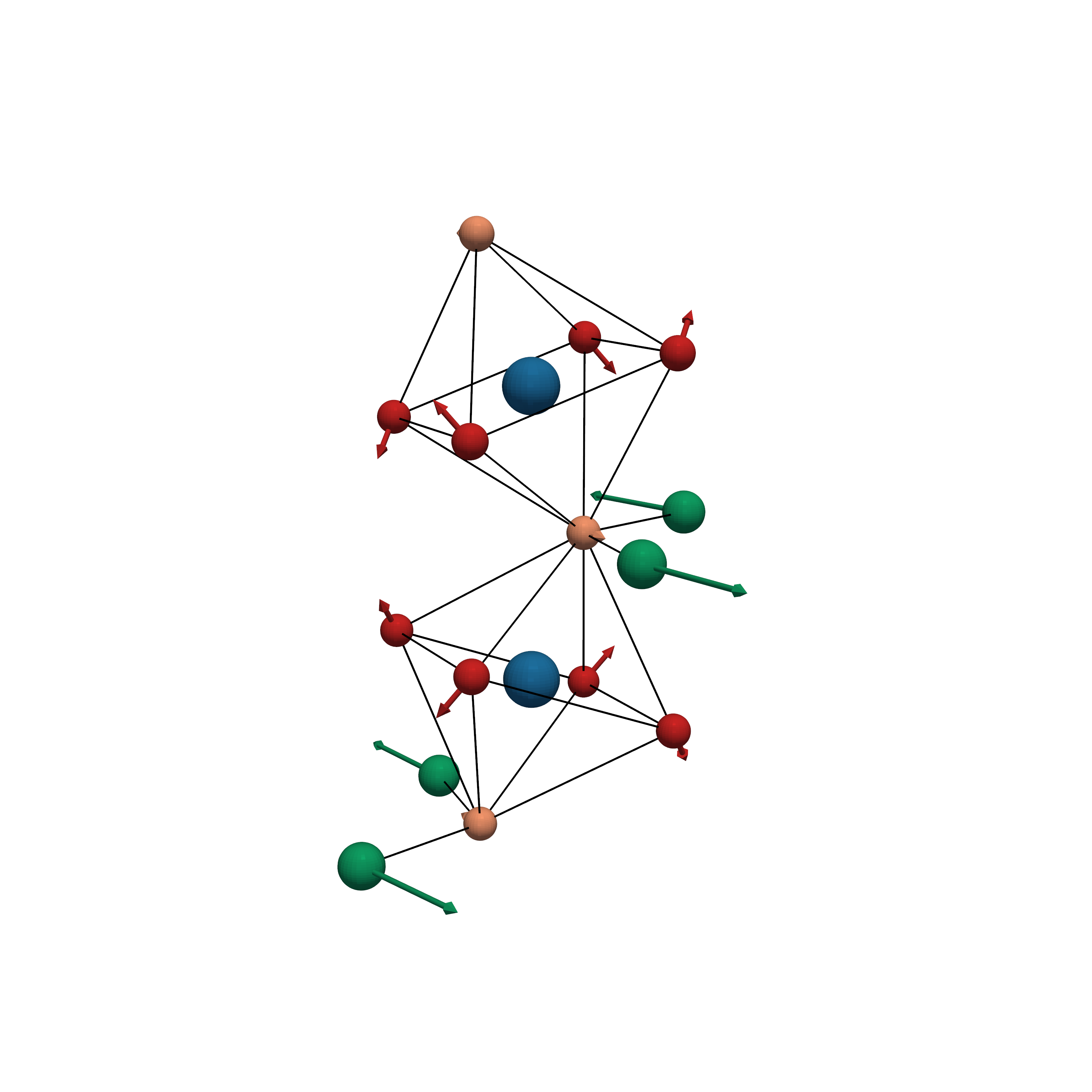}
    \put(28,84){\sffamily\Large $\rm{B^2_{2g}}$}
\end{overpic}
\hfill
\begin{overpic}[width=0.24\linewidth,
                trim=25cm 10cm 30cm 5cm, clip]
               {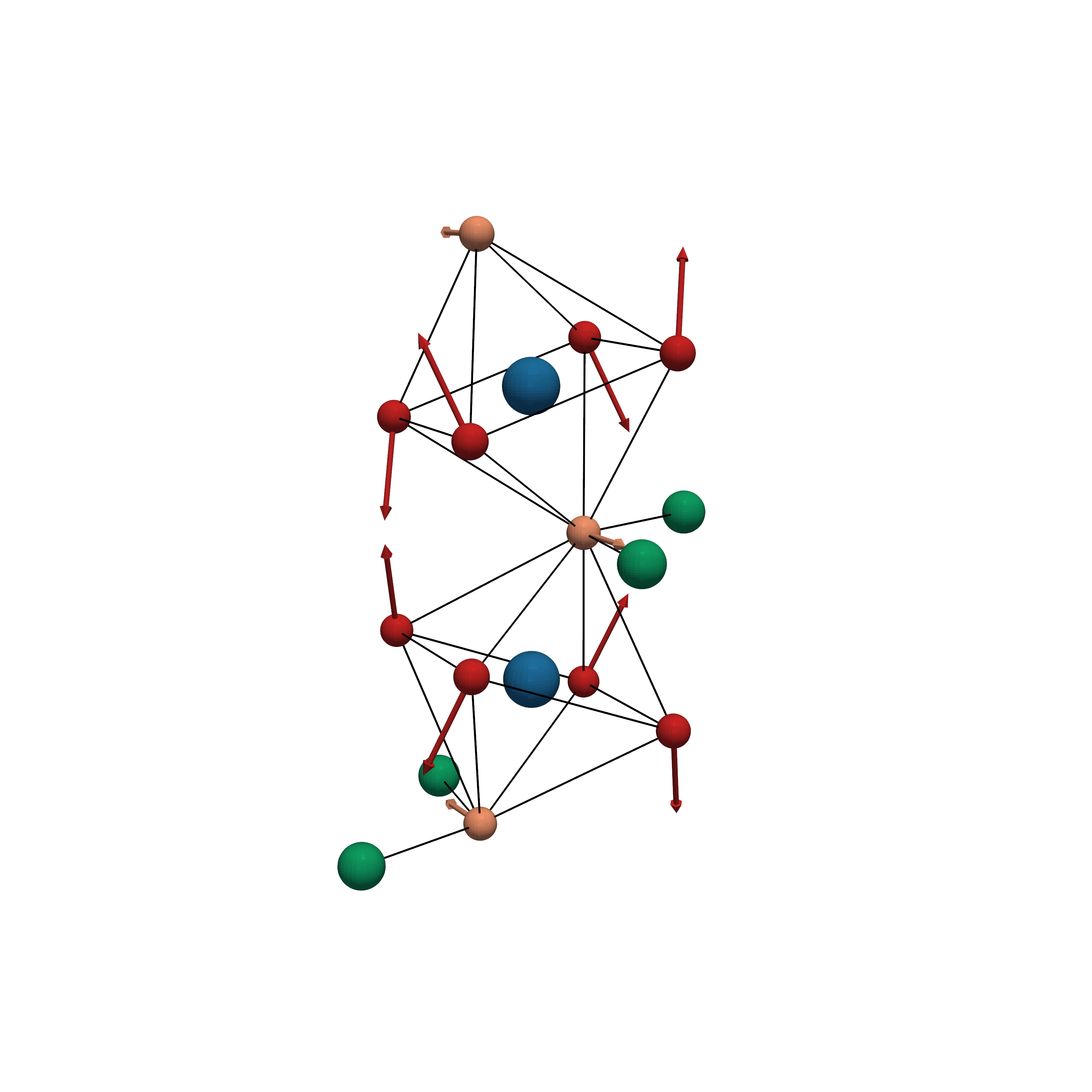}
    \put(28,84){\sffamily\Large $\rm{B^3_{2g}}$}
\end{overpic}
\hfill
\begin{overpic}[width=0.24\linewidth,
                trim=25cm 10cm 30cm 5cm, clip]
               {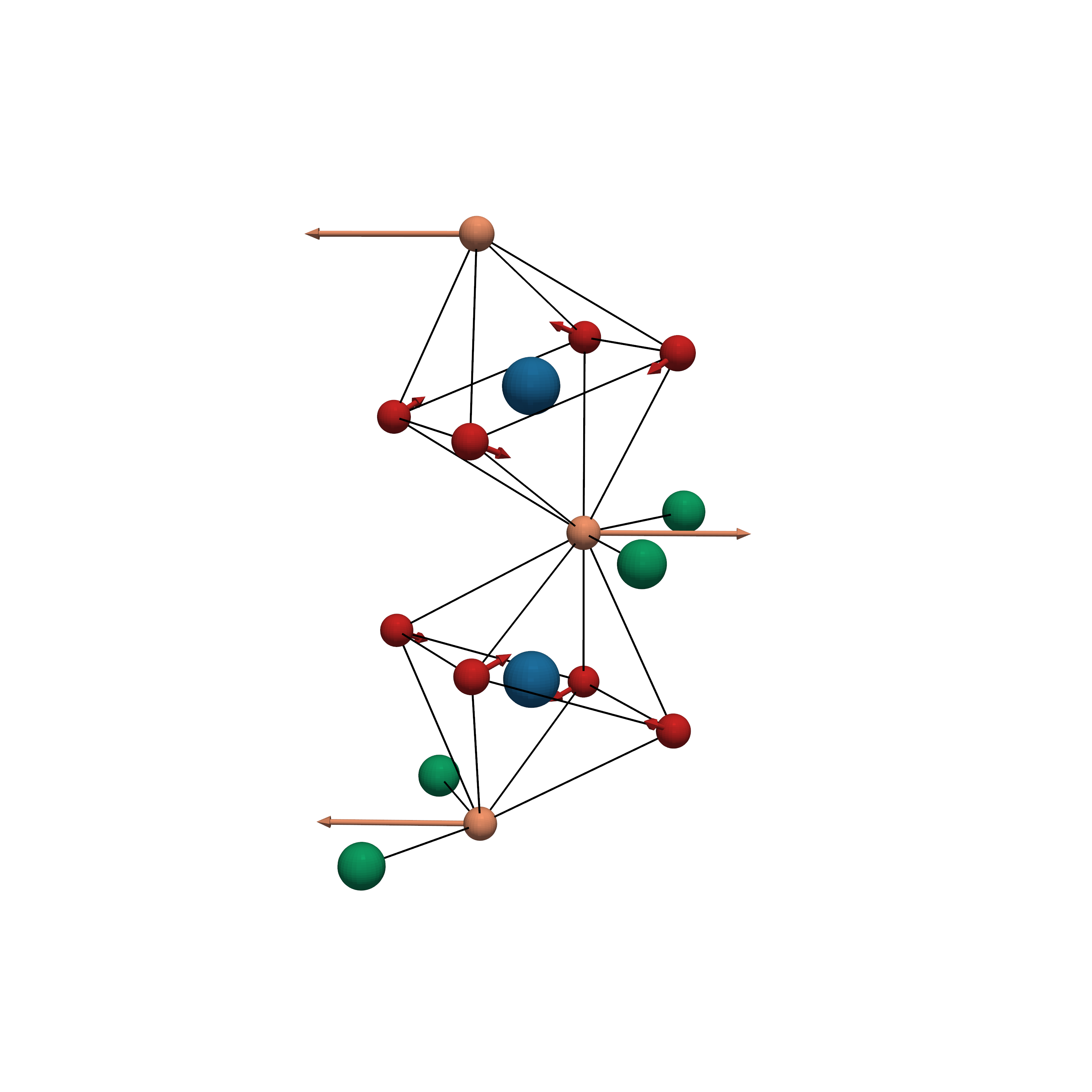}
    \put(28,84){\sffamily\Large $\rm{B^4_{2g}}$}
\end{overpic}
\hfill \\

\begin{overpic}[width=0.24\linewidth,
                trim=25cm 10cm 30cm 5cm, clip]
               {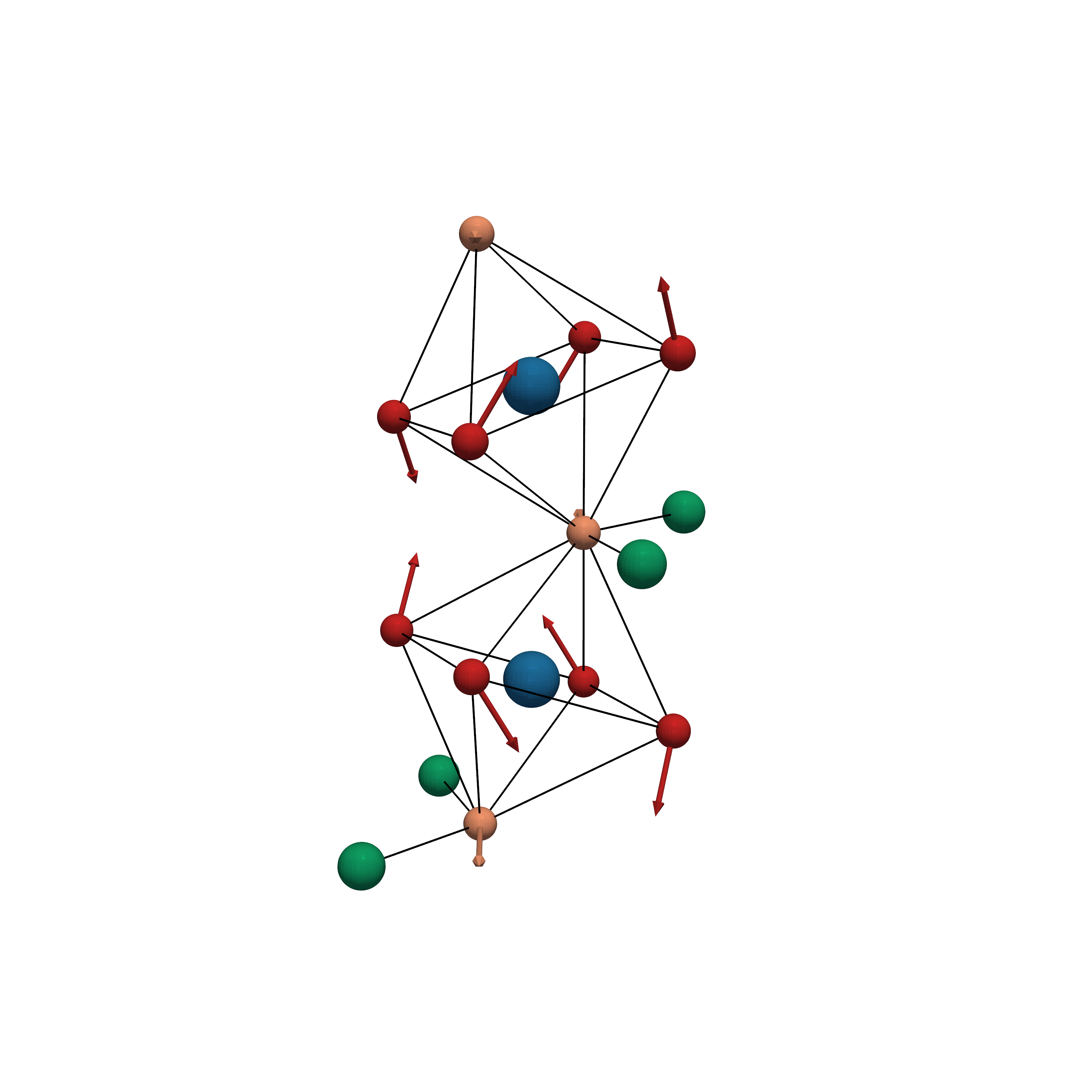}
    \put(28,84){\sffamily\Large $\rm{B^5_{2g}}$}
\end{overpic}
\hfill
\begin{overpic}[width=0.24\linewidth,
                trim=25cm 10cm 30cm 5cm, clip]
               {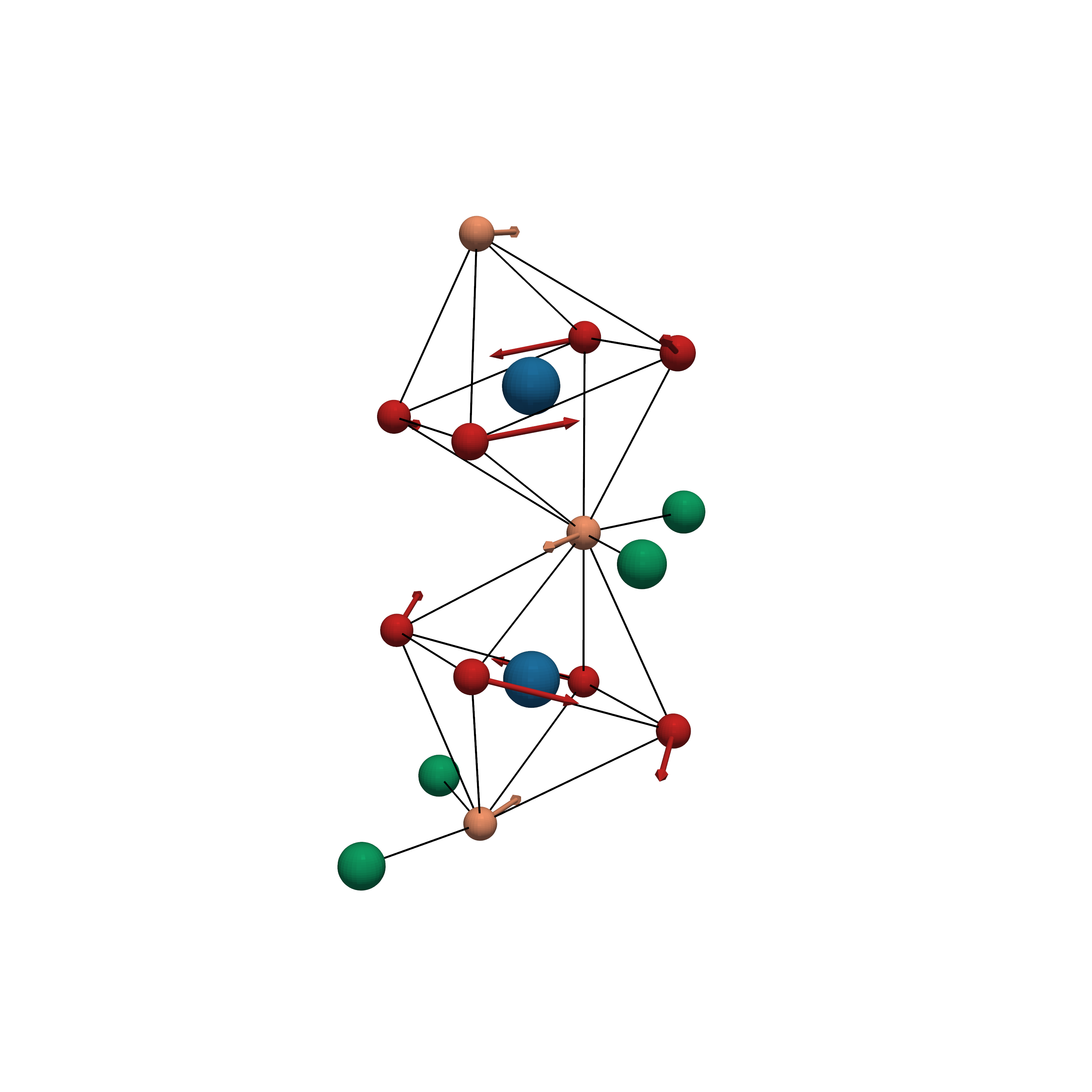}
    \put(28,84){\sffamily\Large $\rm{B^6_{2g}}$}
\end{overpic}
\hfill
\begin{overpic}[width=0.24\linewidth,
                trim=25cm 10cm 30cm 5cm, clip]
               {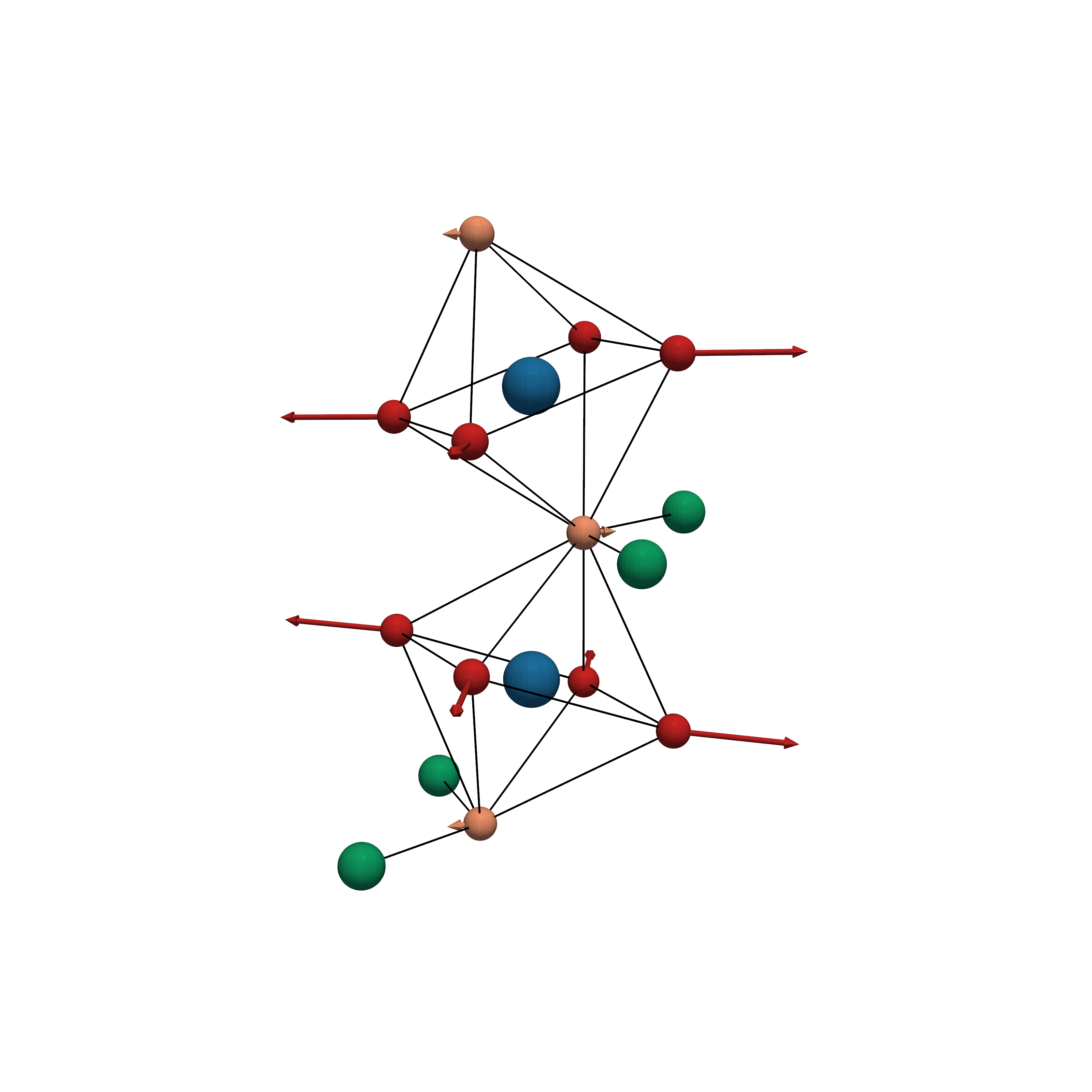}
    \put(28,84){\sffamily\Large $\rm{B^7_{2g}}$}
\end{overpic}
\hfill
\begin{overpic}[width=0.238\linewidth,
                trim=25cm 10cm 30cm 5cm, clip]
               {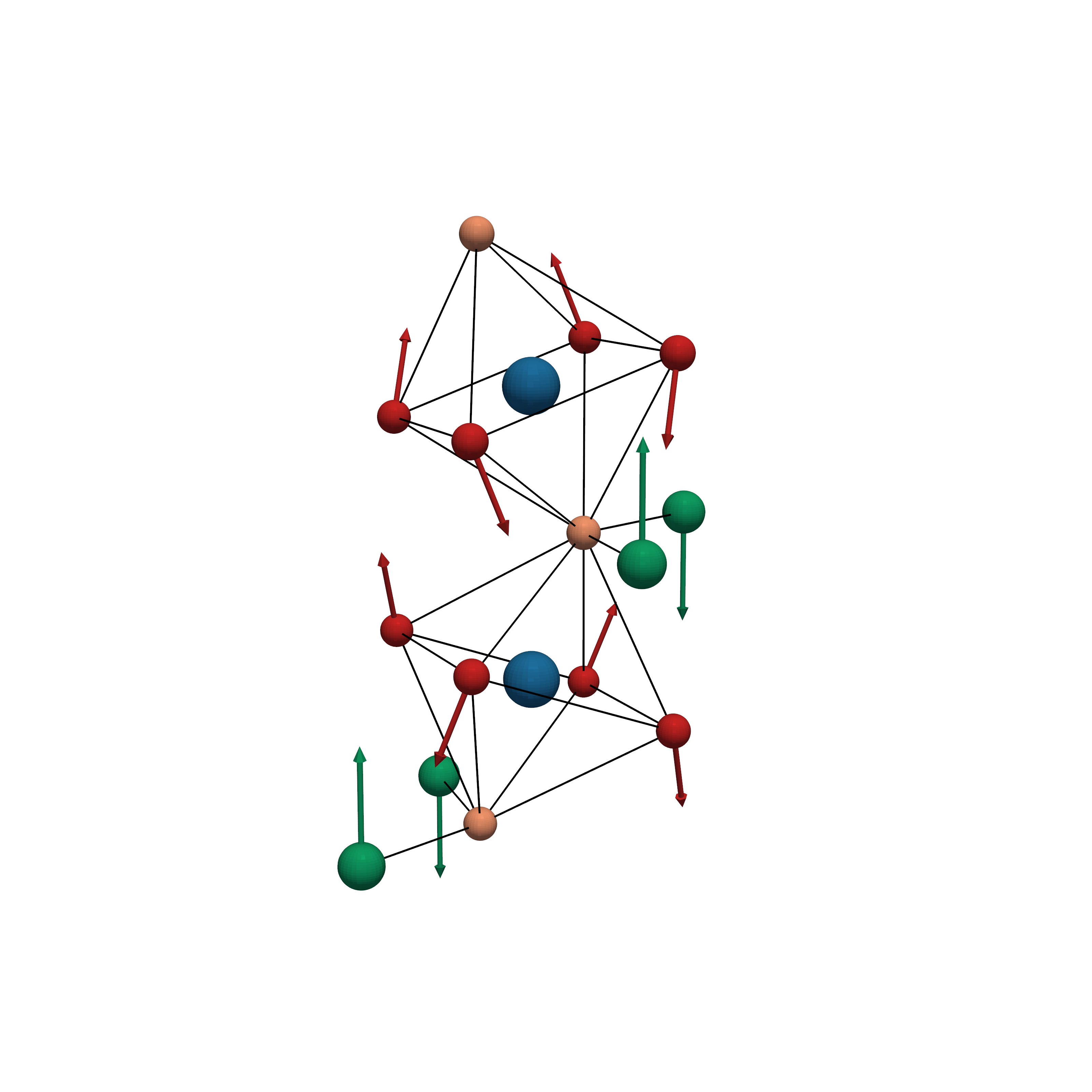}
    \put(28,84){\sffamily\Large $\rm{B^1_{3g}}$}
\end{overpic}
\hfill 

\begin{figure}[H]
\begin{overpic}[width=0.238\linewidth,
                trim=25cm 10cm 30cm 5cm, clip]
               {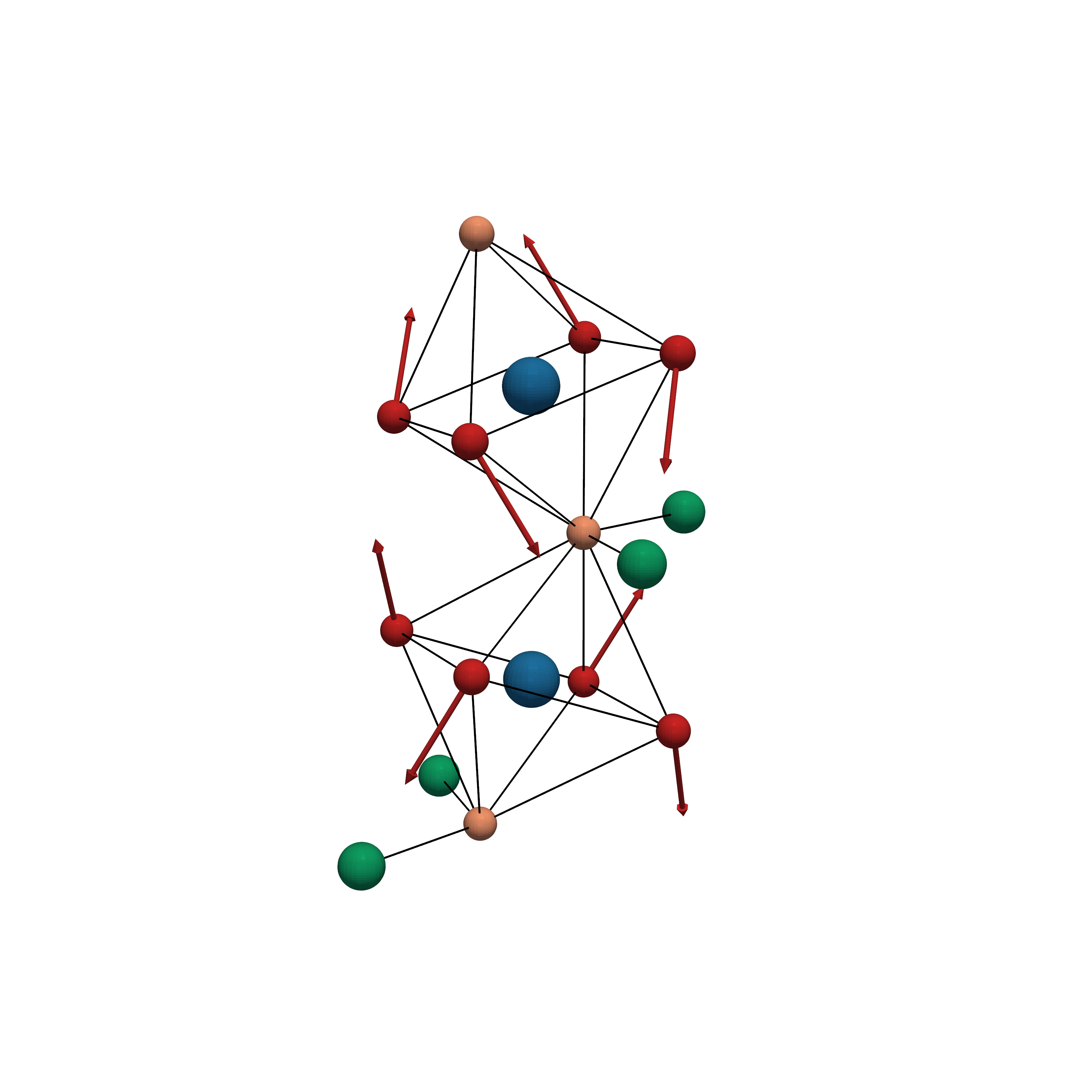}
    \put(28,84){\sffamily\Large $\rm{B^2_{3g}}$}
\end{overpic}
\hfill
\begin{overpic}[width=0.238\linewidth,
                trim=25cm 10cm 30cm 5cm, clip]
               {modes/B3g_3.png}
    \put(28,84){\sffamily\Large $\rm{B^3_{3g}}$}
\end{overpic}
\hfill
\begin{overpic}[width=0.238\linewidth,
                trim=25cm 10cm 30cm 5cm, clip]
               {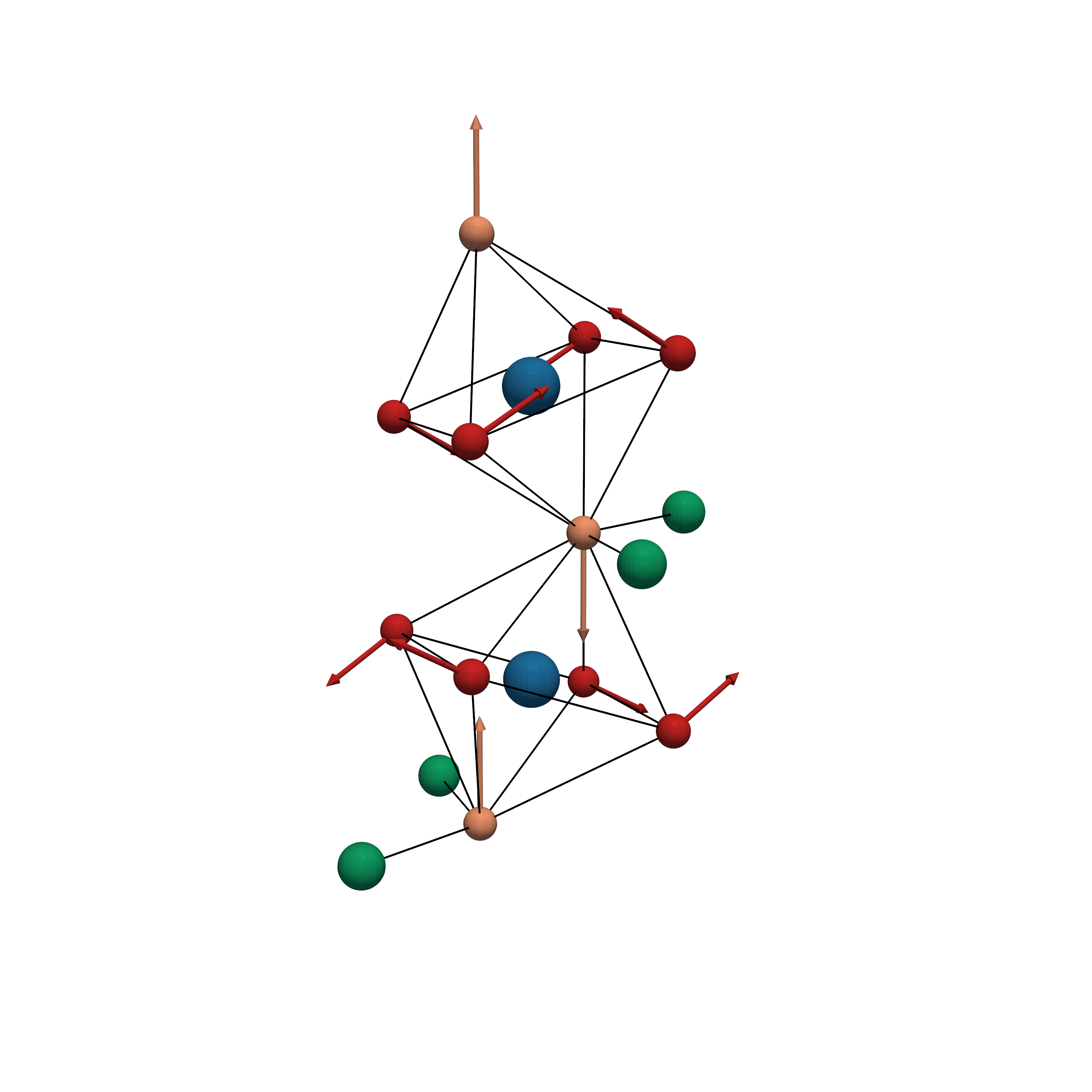}
    \put(28,84){\sffamily\Large $\rm{B^4_{3g}}$}
\end{overpic}
\hfill
\begin{overpic}[width=0.238\linewidth,
                trim=25cm 10cm 30cm 5cm, clip]
               {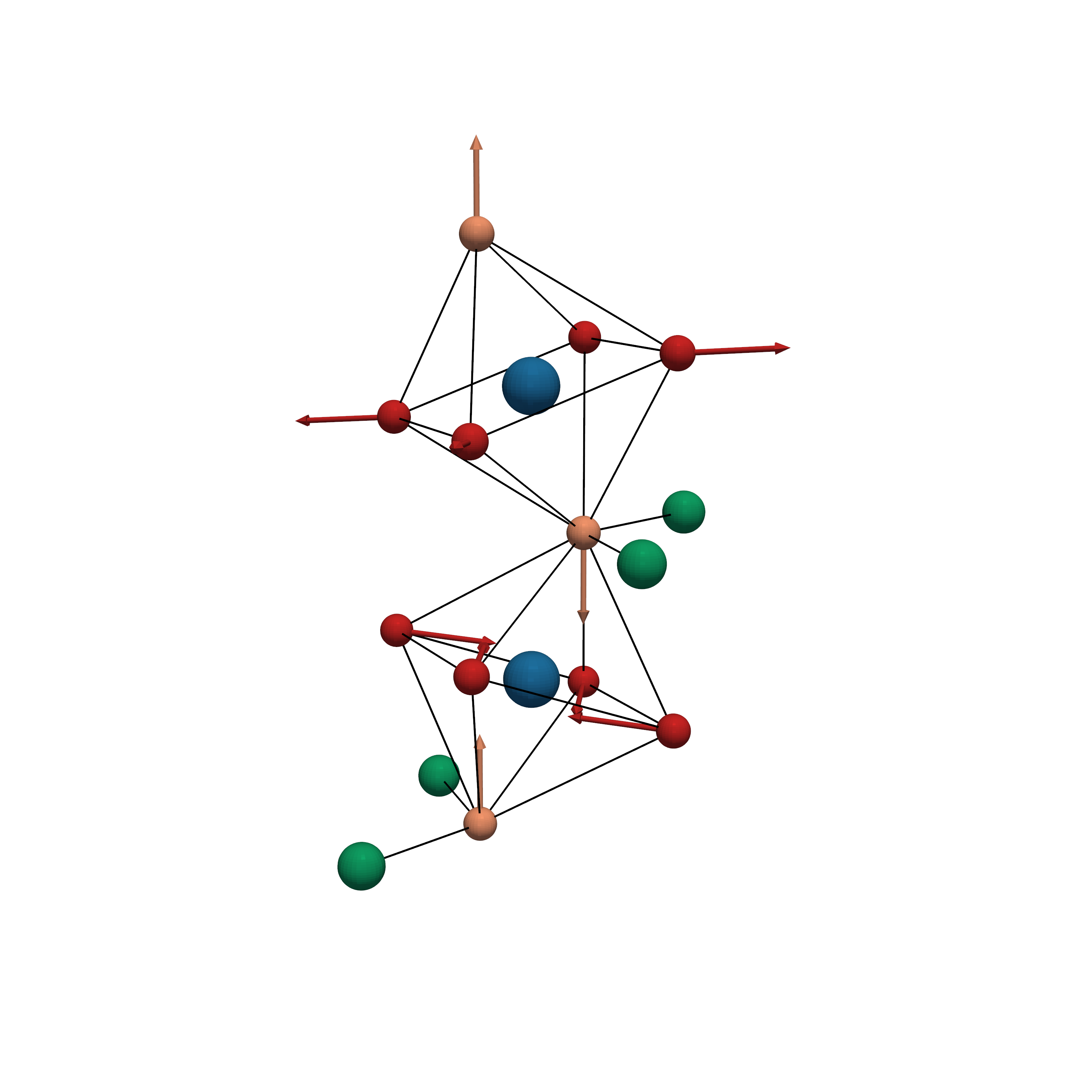}
    \put(28,84){\sffamily\Large $\rm{B^5_{3g}}$}
\end{overpic}
\caption{Phonon displacement patterns of all 24 Raman active modes in LIO. La-, In-, O(1) and O(2)-atoms are depicted in green, blue, orange and red, respectively.}
\label{fig:allemoden}
\end{figure}
\hfill

\begin{figure}
    \centering
    \includegraphics[width=0.7\linewidth]{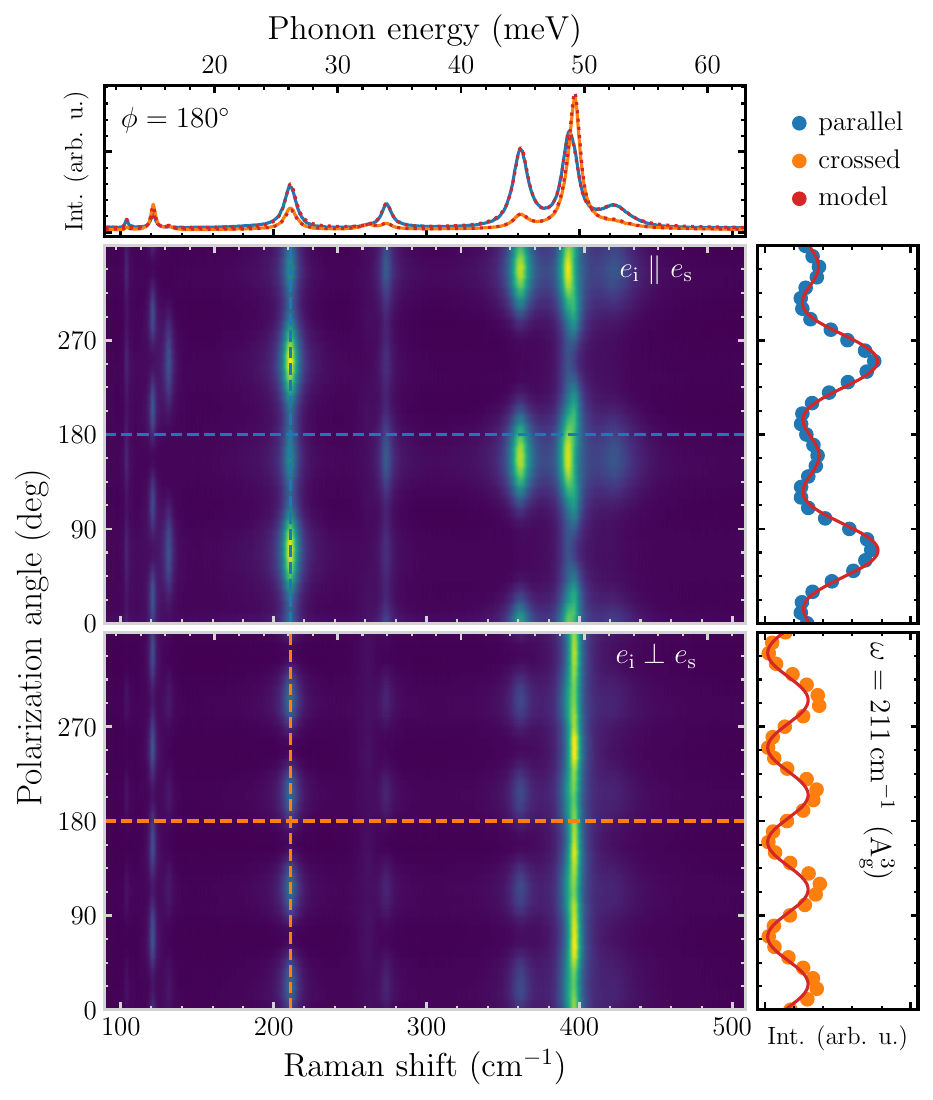}
    \caption{Polarization-angle resolved Raman spectra of the a plane analogously to Fig.\ 2.}
\end{figure}

\begin{figure}
    \centering
    \includegraphics[width=0.7\linewidth]{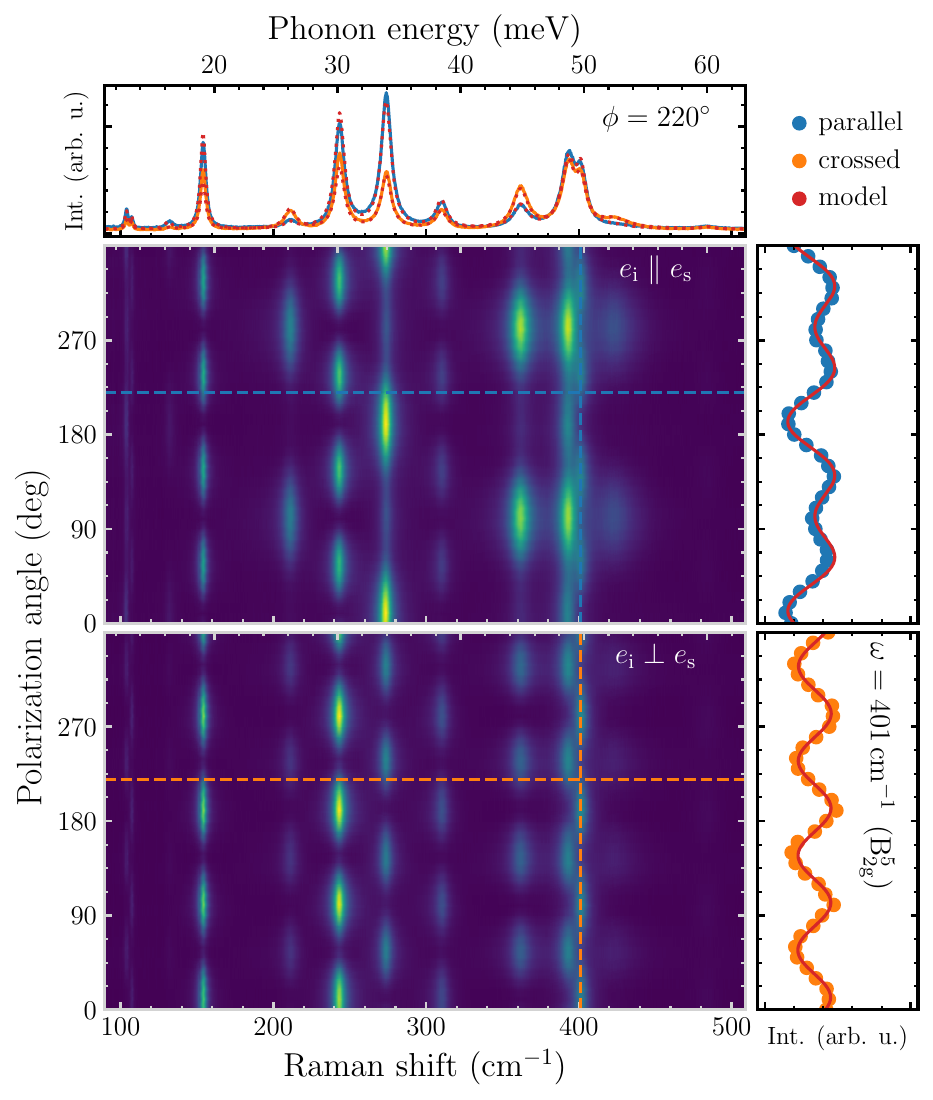}
    \caption{Polarization-angle resolved Raman spectra of the b plane analogously to Fig.\ 2. The intensity slice of the $\rm{B^5_{2g}}$ mode at \SI{401}{\per\centi\metre} (right panels) contains contributions from the $\rm{A^6_g}$ mode at \SI{393}{\per\centi\metre} but is perfectly described within our model.}
\end{figure}

\begin{figure}
    \centering
    \includegraphics[width=0.7\linewidth]{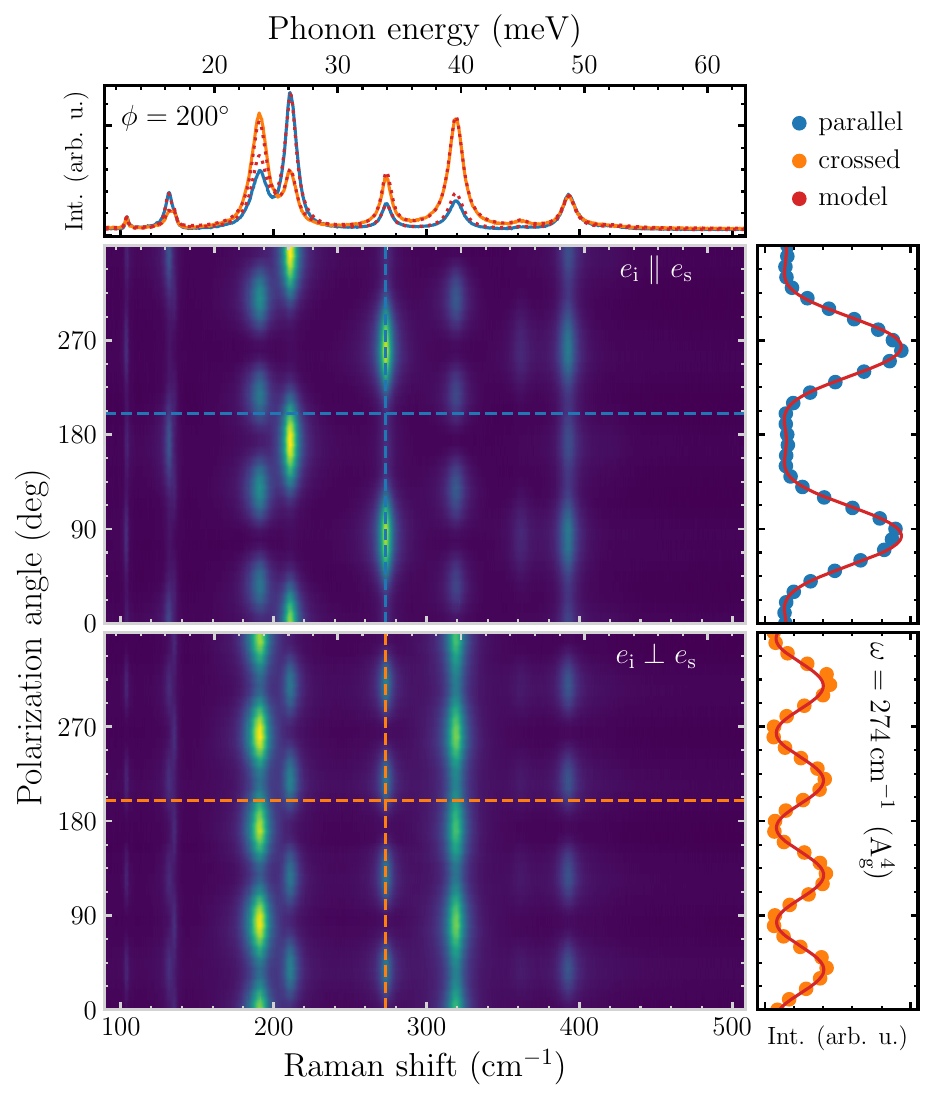}
    \caption{Polarization-angle resolved Raman spectra of the c plane analogously to Fig.\ 2.}
\end{figure}